\documentclass[preprint,notoc]{JHEP3}
\usepackage{epsfig}

\def\eslt{E_T^{\rm miss}}

\def\to{\rightarrow}

\def\bi{\begin{itemize}}
 \def\ei{\end{itemize}}
\def\te{\tilde e}

\def\c1p{C1^\prime}
\def\msq3{\overline{m}_{\tilde{q}}(3)}
\def\ta{\tilde a}

\def\tu{\tilde u}

\def\ta{\tilde a}

\def\tb{\tilde b}

\def\tst{\tilde t}
\def\ttau{\tilde \tau}
\def\tmu{\tilde \mu}
\def\tg{\tilde g}
\def\tnu{\tilde\nu}
\def\tell{\tilde\ell}
\def\tq{\tilde q}

\def\tw{\widetilde W}
\def\tz{\widetilde Z}
\def\alt{\stackrel{<}{\sim}}
\def\agt{\stackrel{>}{\sim}}
\def\be{\begin{equation}}  
\def\ee{\end{equation}}  
\def\bea{\begin{eqnarray}}  
\def\eea{\end{eqnarray}}  

\def\sps1ap{SPS1a$^\prime$}

\title{Natural Supersymmetry: \\
LHC, dark matter and ILC searches
}
\author{Howard Baer$^{a}$, Vernon Barger$^b$, Peisi Huang$^b$ 
and Xerxes Tata$^c$\\
$^a$Dept.\ of Physics and Astronomy, University of Oklahoma, Norman, OK 73019, USA\\
$^b$Dept. of Physics, University of Wisconsin, Madison, WI 53706, USA\\
$^c$Dept.\ of Physics and Astronomy, University of Hawaii, Honolulu, HI 96822, USA\\
E-mail: \email{baer@nhn.ou.edu}, \email{barger@pheno.wisc.edu},
\email{phuang7@wisc.edu}, \email{tata@phys.hawaii.edu}}

\preprint{\vbox{UH-511-1190-12}}

\abstract{Particle physics models with {\it Natural Supersymmetry} 
are characterized by a superpotential parameter $\mu\sim m_h\sim 125$~ GeV, 
while third generation squarks have mass $\alt 0.5-1.5$ TeV. Gluinos
should be lighter than several TeV so as not to destabilize the lighter
squarks. 
First and second generation sfermions can be at the tens-of-TeV level 
which yields a decoupling solution to the SUSY flavor and $CP$ problems. 
Adopting a top-down approach, we delineate the range
of GUT scale SUSY model parameters which leads to a 
Natural SUSY mass spectrum. 
We find natural SUSY models to be tightly constrained by the 
$b\to s\gamma$ branching fraction measurement while it is also difficult but not impossible to
accommodate a light Higgs scalar of mass $\simeq 125$ GeV.
We present several benchmark points which are expandable to slopes and planes.
Natural SUSY is difficult to see at LHC unless some third generation squarks are
very light. The top- and bottom- squarks cascade decay mainly to higgsino-like 
charginos and neutralinos via numerous posibilities, 
leading to a rather complex set of signatures. 
Meanwhile, a linear $e^+e^-$ collider operating at $\sqrt{s}\sim 0.25-0.5$ TeV 
would be a {\it higgsino factory} and is essentially guaranteed a SUSY discovery 
of the low-lying charged and neutral higgsino states.
Since thermal neutralino cold dark matter is underproduced, we conjecture that 
the incorporation of a Peccei-Quinn sector or light moduli into the theory will augment higgsino
dark matter production, possibly together with an admixture of axions.
We present rates for direct and indirect higgsino dark matter detection for the case
where light higgsinos dominate the dark matter abundance.
}
\keywords{Supersymmetry Phenomenology, Supersymmetric Standard Model,
Large Hadron Collider}

\begin{document}

\section{Introduction}
\label{sec:intro}

\subsection{Impact of LHC sparticle searches}

The search for weak scale supersymmetry (SUSY) \cite{susy,wss,rev} has
begun in earnest at the CERN Large Hadron Collider (LHC).
From a non-observation of multi-jet plus multi-lepton + $\eslt$ events
with or without tagged $b$-jets in a data sample of
$\sim1-4.5$~fb$^{-1}$, the CMS\cite{cms} and ATLAS \cite{atlas}
experiments have excluded gluinos and squarks up to 1.4~TeV for
$m_{\tg}\simeq m_{\tq}$ and gluinos up to $\sim 0.8$~TeV, for the
case of $m_{\tq}\gg m_{\tg}$\cite{lowette}.

Many experimental analyses have been performed within the framework of
the mSUGRA (or CMSSM) model, which assumes a common mass parameter $m_0$
(renormalized at the GUT scale) for all scalars, and likewise a common
mass parameter $m_{1/2}$ for the gauginos.  The physical spectrum--
obtained by renormalization group (RG) running of soft mass parameters
from $M_{\rm GUT}$ to $M_{\rm weak}$-- is characterized by a squark mass
spectrum with $m_{\tq}\sim m_{\tg}$ for low $m_0 \alt m_{1/2}$ or
$m_{\tq}\gg m_{\tg}$ for large values of $m_0$.  Despite the fact that
superpotential Yukawa interactions reduce third generation squark masses
relative to those of first/second generation squarks, third generation
squarks nonetheless frequently have masses $\agt 1$ TeV, putting them in
conflict with electroweak fine-tuning constraints (discussed below).
This has led some physicists to question whether weak scale SUSY indeed
stabilizes the gauge hierarchy, given the constraints from the
LHC.\footnote{By adjusting the trilinear soft breaking parameter $A_0$
to certain values, the $\tst_1$ mass may be dialed to sub-TeV
values. However, the remaining third generation squark masses typically
remain at large values and still in possible conflict with fine-tuning
constraints.}

We emphasize that while the various squarks end up being nearly
degenerate within the mSUGRA model, the limit on $m_{\tq}$ quoted above
arises mainly from the production and decay of {\it first-generation}
squarks. Only these squarks can be pair produced from the valence $u$
and $d$ quark content of the colliding protons. As a result, their
production cross section falls off much less rapidly with increasing
squark mass than the corresponding cross section for the production of
second and third generation squarks: thus, the ATLAS and CMS limits,
$m_{\tq}\agt 1$ TeV, apply to first generation squarks, while second and
third generation squarks may be much lighter without being in conflict
with either LHC data or with the notion of superpartners as the new physics that
stabilizes the weak scale.

\subsection{Impact of LHC Higgs searches}
\label{ssec:h125}

Recent results from LHC Higgs searches find tantalizing hints for a
Standard Model (SM)-like Higgs boson of mass $m_h\simeq 125$ GeV
(although at present values of $m_h\sim 120$ GeV are also
possible). Such a large value of $m_h$ is difficult to realize in models
such as minimal anomaly mediation (mAMSB) or minimal gauge mediation
(mGMSB)\cite{djouadi,dm125} unless {\em all} sparticle masses are in the
10-20 TeV range, in severe conflict with electroweak fine-tuning constraints.
Meanwhile, gravity mediation (SUGRA) remains a possible venue for
communication of SUSY breaking since, unlike in mAMSB and mGMSB models,
the scalar trilinear soft SUSY breaking coupling $A_0$ is an independent
parameter, and can be chosen to be large as seems to be required by such
large values of $m_h$\cite{h125}. 

SUSY models based on gravity-mediation are, however, expected
generically to give rise to large FCNC and CP-violating
processes\cite{kl} since there is no mechanism to enforce the required
generational universality\cite{susy,wss,rev} or alignment of fermion and
sfermion mass matrices needed to reduce flavour-changing processes to an
acceptable level.\footnote{In mSUGRA, the SUSY GIM mechanism is imposed
by simply assuming universality of scalar masses at the high scale,
usually taken to be $M_{\rm GUT}$.}  Indeed, the SUSY flavor and CP
problem endemic to gravity-mediation has served as motivation for the
construction of AMSB and GMSB models, since SUSY sources of FCNC and
$CP$-violation are automatically suppressed in these models.

An alternative solution to the SUSY flavor and $CP$ problems arises by
{\it decoupling}: allowing for first and second generation squark and
slepton masses to be in the 10-50 TeV range. Third generation squark masses,
which directly enter into electroweak fine-tuning, or ``naturalness''
considerations (see below), may be much lighter since flavor and $CP$
constraints are relatively mild for third generation
particles\cite{masiero}.  Supersymmetric models containing a split
spectrum -- sub-TeV third generation squarks but with multi-TeV
first/second generation squarks -- have been advocated for some time under
the label of {\it effective SUSY}\cite{ckn}, or
ESUSY\cite{esusy}. Indeed, the non-observation of squarks and gluinos in
the LHC data sample could be a hint pointing in this direction.

\subsection{Naturalness constraints}
\label{ssec:nat}

It is well known \cite{wss} that at tree-level the magnitude of the
higgsino mass parameter $\mu$ is determined in terms of (1)~the weak
scale soft SUSY breaking (SSB) mass parameters $m_{H_u}^2$ and
$m_{H_d}^2$ that appear in the Higgs sector scalar potential, (2)~the
ratio $\tan\beta\equiv \frac{v_u}{v_d}$, and (3)~the observed value of the
$Z$-boson mass.  Including radiative corrections via the effective
potential method, this relation gets modified to:
\begin{equation}
 \frac{1}{2}M_Z^2
=\frac{(m_{H_d}^2+\Sigma_d)-(m_{H_u}^2+\Sigma_u)\tan^2\beta}{(\tan^2\beta
  -1)} - \mu^2\;. 
\label{eq:zmass}
\end{equation}
Here, $\Sigma_u$ and $\Sigma_d$ arise from radiative corrections\cite{anradcorr}, 
and are given in the 1-loop approximation to the Higgs effective potential by
$$\Sigma_{u,d}= \frac{1}{v_{u,d}}\frac{\partial \Delta V}{\partial H_{u,d}},$$ 
where $\Delta V$ is the one-loop correction to the tree-level potential,
and the derivative is evaluated in the physical vaccuum: {\it i.e.} the
fields are set to their vacuum expectation values after evaluating the
derivative.
  
It is reasonable to say that the theory {\it naturally} yields the correct value of $M_Z$
if the individual terms on the right hand side of Eq.~(\ref{eq:zmass})
are comparable in magnitude so that the observed value of $M_Z$ is
obtained without resorting to large cancellations. Indeed this is why
$|\mu|$ has been suggested as a measure of naturalness\cite{nath-chat},
with theories where $\mu^2 \alt M_Z^2$ being the ``most natural''. 
This relationship must be accepted with some latitude, since values of 
$\mu^2 \alt$ (100~GeV)$^2$ are phenomenologically excluded by the LEP2 limit
that $m_{\tw_1}>103.5$ GeV. 
Of course, there is nothing special about $\mu^2$ and the same  considerations
apply equally to all the terms, including those involving the radiative corrections. 

In the following, we will somewhat arbitrarily require that each
individual term in (\ref{eq:zmass}) is bounded by about (200~GeV)$^2$.
Similar considerations have recently been adopted by several other
groups\cite{arkani,sundrum,essig,papucci,nevzorov}.\footnote{These
analyses differ in detail on the restrictions on each term, and even
whether a common constraint is applied to each term. For this reason,
some of the constraints that have been obtained by these analyses are
stronger than the ones we obtain in this paper.}  In distinction with
other works, our focus is on the expected sparticle mass spectra and
collider and dark matter phenomenology of Natural SUSY models with
parameters defined at a high scale (taken to be $M_{GUT}$) which lead to
weak scale parameters that are natural in the sense that we have just
described.

The largest contributions to $\Sigma_{u,d}$ in Eq.~(\ref{eq:zmass})
arise from superpotential  Yukawa interactions of third generation
squarks involving the top quark Yukawa coupling. 
The dominant contribution to these quantities is given by
$$\Sigma_u \sim \frac{3f_t^2}{16\pi^2}\times m_{\tst_i}^2 \left(\ln
(m_{\tst_i^2}/Q^2) -1\right)\;,$$ 
and so grows quadratically with the top squark masses. 
Clearly, the top squark (and by $SU(2)$ gauge symmetry, also $\tb_L$) masses must then be
bounded above by the naturalness conditions. In Ref.~\cite{nevzorov}, it
has been shown that requiring  $\Sigma_u \alt \frac{1}{2}M_Z^2$ leads to
$m_{\tst_i}\alt 500$~GeV. Scaling this to allow $\mu$ values up to 150~GeV
(200~GeV) leads to a corresponding bound $m_{\tst_i} \alt 1$~TeV (1.5~TeV), which
of course also applies to the {\it heavier} top 
squark.\footnote{For very large values of $\tan\beta$ where the bottom quark Yukawa becomes 
comparable to $f_t$, similar considerations would apply to bottom squarks.} 
In other words, from this perspective, models with $\mu \alt 150-200$~GeV and top squarks at the
TeV scale or below are completely natural. 
In this connection, it is perhaps worth remarking that since 
\begin{equation}
m_A^2 \simeq 2\mu^2 + m_{H_u}^2 + m_{H_d}^2 + \Sigma_u + \Sigma_d\;,
\label{eq:Amass}
\end{equation}
for moderate to large values of $\tan\beta$, 
the heavier Higgs scalars can naturally be at the several-TeV scale because of the appearance of
$\tan^2\beta-1$ in the denominator of Eq.~(\ref{eq:zmass}). 
Notice, however, that the bound of (200 GeV)$^2$ on each term in
Eq.~(\ref{eq:zmass}) translates to an upper bound 
\be
m_A\sim \left|m_{H_d}^2\right|^{1\over 2} \alt |\mu | \tan\beta \ .
\ee
Such a constraint could prove theoretically significant in considerations of 
high scale models with special properties such as models with unified Yukawa couplings 
at the GUT scale\cite{yukawa}.

Our discussion up to this point shows that SUSY models with $|\mu| \alt
150-200$~GeV and top squark masses (and if $\tan\beta$ is very large, also bottom squark masses) 
below 1-1.5~TeV are perfectly natural. There will also be
corresponding constraints on other sparticles such as electro-weak
charginos and neutralinos that also directly couple to the Higgs sector,
but since these couplings are smaller than $f_t$ and because there are no
colour factors, the constraints will be correspondingly weaker.
Sparticles such as first and second generation squarks and sleptons that
have no direct/significant couplings to the Higgs sector are constrained
only via two-loop effects and can easily be in the 10 - 50 TeV range.
An important exception would be the gluino, since radiative corrections
to the top squark mass are proportional to $m_{\tg}$\cite{sundrum}.
Using $\delta m_{\tq}^2 \sim \frac{2g_s^2}{3\pi^2}m_{\tg}^2\times log$
and setting logs to be order unity, we see that $m_{\tg} \alt
3m_{\tq}$. For top squarks to remain below the 1.5~TeV range, the gluino
must be lighter than about 4~TeV. In models with electroweak gaugino mass
unification, electroweak-inos would then automatically not destroy
naturalness.

\subsection{Natural SUSY}
\label{ssec:ns}

These considerations suggest that the region of SUSY parameter space where
\bi
\item $|\mu |\alt 150-200$~GeV, 
\item third generation squarks $m_{\tst_{L,R}},\ m_{\tb_L}\alt 1-1.5$~TeV,
\item $m_{\tg} \alt 3-4$~TeV and SSB electroweak-ino masses smaller than 1-2~TeV
\item $m_A\alt |\mu |\tan\beta$,
\item $m_{\tq_{1,2}},\ m_{\tell_{1,2}}\sim 10-50$ TeV, \ei 
may, from
naturalness and flavor/$CP$ considerations, merit a dedicated study.  The first
and second generation squarks and sleptons -- lying in the 10-50 TeV
range -- provide a decoupling solution to the SUSY flavour problem, the
SUSY $CP$ problem and to the problem of too-rapid-proton decay. We
remark here that if SUSY breaking arises from supergravity breaking in a
hidden sector, then the gravitino mass $m_{3/2}$ sets the scale for the
largest of the SSB terms, and we would also expect $m_{3/2}\sim 10-50$
TeV: such a high value of $m_{3/2}$ also provides a solution to the gravitino
problem\cite{linde,moroi}.  The heavier Higgs bosons may easily be in
the several-TeV range for moderate to large values of $\tan\beta$.

SUSY models with the above generic spectra have been dubbed ``Natural
SUSY''\cite{essig} and are a more restrictive case of effective SUSY
models because we further restrict $|\mu |\alt 150-200$ GeV.
This usually gives rise to a higgsino-like lightest neutralino $\tz_1$.

\subsection{Dark matter in natural SUSY}

In fact, a problem with effective SUSY models with a bino-like $\tz_1$
arises in that a vast overabundance of neutralino cold dark matter (CDM)
is expected\cite{esusy}, typically 2-4 orders of magnitude above the
WMAP-measured value of $\Omega_{CDM}h^2\sim 0.11$ unless weak scale
parameters happen to be in special parameter space regions.
It has been suggested in \cite{esusy} that if the strong CP problem is
solved by the Peccei-Quinn (PQ) mechanism -- which introduces a
supermultiplet containing spin-zero axion and saxion fields, along with
a spin-${1\over 2}$ axino -- then neutralinos might decay to a light
axino LSP via $\tz_1\to\ta\gamma$. Since each neutralino converts to one
axino, the decay-produced axino abundance is given by
$\Omega_{\ta}^{NTP}=m_{\ta}/m_{\tz_1}\Omega_{\tz_1}h^2$. For $m_{\ta}$
in the MeV range, the suppression factor is $\sim 10^{-3}-10^{-5}$,
bringing the DM abundance into accord with measurement.

However, typically in gravity mediation models the axino mass is
expected to be around the TeV scale\cite{cl,hall}, with $\tz_1$
remaining as LSP. In fact, in the PQ-augmented SUSY model, one then
expects thermal axino production (TP) in the early universe, followed by
late-time $\ta\to \tz_1\gamma$ decays, so the dark matter overabundance
is made even worse. In the case of natural SUSY, the higgsino-like
$\tz_1$ leads typically to a thermal {\it underabundance} of neutralino
CDM. But now TP axinos followed by their decay to neutralinos can {\it
augment} this abundance\cite{ckls,blrs,bls}, while any remaining
underabundance can be filled by axions produced via vacuum mis-alignment
(coherent oscillations).  Thus, in this case we might expect the CDM to
consist of a higgsino-axion admixture.  Which of these two particles
dominates the DM abundance depends on specific choices of PQ parameters
and on the value of re-heating temperature $T_R$ after inflation.

\section{Parameter space and mass spectra for Natural SUSY}
\label{sec:pspace}

Since the introduction of softly broken SUSY into the Standard Model
(leading to the Minimal Supersymmetric Standard Model, or MSSM) leads to
a theory with stable mass hierarchies, it is natural to assume the MSSM
is the low energy effective theory arising from an underlying SUSY Grand
Unified  
Theory (GUT) which is broken at some high energy scale, taken here
for definiteness to be $M_{GUT}\simeq 2\times 10^{16}$ GeV.  Indeed, the
MSSM (or MSSM plus gauge singlets and/or additional complete $SU(5)$
multiplets) receives some indirect support from experiment in that
1. the measured weak scale gauge couplings unify nearly to a point at
$M_{GUT}$ under MSSM renormalization group (RG) evolution and 2. the
MSSM electroweak symmetry is broken radiatively due to the large top
quark Yukawa coupling, consistent with the measured value of $m_t$.

Motivated by these SUSY success stories, the interesting question arises
as to whether the natural SUSY sparticle mass spectrum can be
consistently generated from a model with parameters defined at the high
scale $Q=M_{GUT}$.  To implement a low value of $|\mu |\alt 150-200$
GeV, we will adopt the 2-parameter non-universal Higgs model
(NUHM2)\cite{nuhm2}, wherein weak scale values of $\mu$ and $m_A$ may be
used as inputs in lieu of GUT scale values of $m_{H_u}^2$ and
$m_{H_d}^2$.  To generate the split first/second versus third generation
scalar mass hierarchy, we will adopt a common GUT scale mass $m_0(3)$
for third generation scalars, and a common GUT scale mass $m_0(1,2)$ for
the first/second generation scalars. The intra-generational mass
universality is well-motivated by $SO(10)$ GUT symmetry, since all
matter multiplets of a single generation live in the 16-dimensional
spinor rep of $SO(10)$. We can also allow some degree of
non-universality between $m_0(1)$ and $m_0(2)$ so long as both lie in
the tens of TeV regime, and provide a decoupling solutions to SUSY FCNC
and $CP$-violating processes (for constraints from FCNC processes, see
Ref. \cite{nmh}). For convenience, we will take them as degenerate.

To allow for a light third generation, we adopt different GUT scale
values for the scalar mass parameter of the first two generations and
the third generation.
In the spirit of SUSY GUT theories, we will assume gaugino mass
unification to a common gaugino mass $m_{1/2}$, and assume a universal
trilinear scalar coupling $A_0$ at the GUT scale.  The sparticle mass
spectrum together with sparticle couplings is then determined by the
parameter set,\footnote{We could have allowed a $D$-term contribution
  that would arise
  when the additional $U(1)$ that is in $SO(10)$ but not in the $SU(5)$
  subgroup is spontaneously broken. However, because we allow $\mu$ and
  $m_A$ as inputs, this would, however, have no impact
  on the allowed range of $m_h$, and so would only affect sparticle
  masses. In order to avoid a time-consuming scan over yet one more parameter
  which would probably not qualitatively alter the main results
  presented below, we have not included this term in our parametrization
  of the model.}
\begin{equation}
m_0(1,2),\ m_0(3),\ m_{1/2},\ A_0,\ \tan\beta,\ \mu,\ m_A \;.
\label{eq:ssb}
\end{equation}
We take $m_t=173.3$ GeV.

Our goal in this Section is to search for weak scale spectra that are
natural in the sense defined above within this framework, to delineate
regions of parameter space consistent with low energy constraints, and
to study their implications for SUSY searches at the LHC.

For our calculations, we adopt the Isajet 7.82~\cite{isajet} SUSY
spectrum generator Isasugra\cite{isasugra}.  Isasugra begins the
calculation of the sparticle mass spectrum with input $\overline{DR}$
gauge couplings and $f_b$, $f_\tau$ Yukawa couplings at the scale
$Q=M_Z$ ($f_t$ running begins at $Q=m_t$) and evolves the 6 couplings up
in energy to scale $Q=M_{GUT}$ (defined as the value $Q$ where
$g_1=g_2$) using two-loop RGEs.  We do not strictly enforce the
unification condition $g_3 = g_1 = g_2$ at $M_{GUT}$, since a few
percent deviation from unification can be attributed to unknown
GUT-scale threshold corrections~\cite{Hisano:1992jj}.  At $Q=M_{GUT}$,
we introduce the SSB parameters in (\ref{eq:ssb}) as boundary
conditions, and evolve the set of 26 coupled MSSM
RGEs~\cite{mv} back down in scale to $Q=M_Z$.  Full two-loop
MSSM RGEs are used for soft term evolution, while the gauge and Yukawa
coupling evolution includes threshold effects in the one-loop
beta-functions, so the gauge and Yukawa couplings transition smoothly
from the MSSM to SM effective theories as different mass thresholds are
passed.  In Isasugra, the values of SSB terms which mix are frozen out
at the scale $Q\equiv M_{SUSY}=\sqrt{m_{\tst_L} m_{\tst_R}}$, while
non-mixing SSB terms are frozen out at their own mass
scale~\cite{isasugra}.  The scalar potential is minimized using the
RG-improved one-loop MSSM effective potential evaluated at an optimized
scale $Q=M_{SUSY}$ which accounts for leading two-loop
effects~\cite{haber}.  Once the tree-level sparticle mass spectrum is
computed, full one-loop radiative corrections are calculated for all
sparticle and Higgs boson masses, including complete one-loop weak scale
threshold corrections for the top, bottom and tau masses at scale
$Q=M_{SUSY}$~\cite{pbmz}.  Since the GUT scale Yukawa couplings are
modified by the threshold corrections, the Isajet RGE solution must be
imposed iteratively with successive up-down running until a convergent
sparticle mass solution is found.  Since Isasugra uses a ``tower of
effective theories'' approach to RG evolution, we expect a more accurate
evaluation of the sparticle mass spectrum for models with split spectra
(this procedure sums the logarithms of potentially large ratios of sparticle masses) 
than with programs which make an all-at-once transition 
from the MSSM to SM effective theories.

We searched for Natural SUSY solutions in the above parameter space by
first fixing $\mu =150$ GeV, and then performing a (linearly weighted)
random scan over the remaining parameters in the following ranges:
\bea
m_0(1,2):\ 5-50\ {\rm TeV},\\
m_0(3):\ 0-5\ {\rm TeV},\\
m_{1/2}:\  0-5\ {\rm TeV},\\
-4<\ A_0/m_0(3)\ <4,\\
m_A:\ 0.15-2\ {\rm TeV},\\
\tan\beta :\ 1-60 .
\label{eq:param}
\eea
We require of our solutions that (1)~electroweak symmetry be radiatively
broken (REWSB), (2)~the neutralino $\tz_1$ is the lightest MSSM
particle, (3)~the light chargino mass obeys the rather model
independent LEP2 limit that $m_{\tw_1}>103.5$ GeV\cite{lep2ino} and
(4)~that $m_{\tg}<4$ TeV, in accord with our naturalness criterion
detailed above.

The results of our scan are plotted in Fig. \ref{fig:param}.  On the
$y$-axis, we plot the average third generation squark mass
\be
\msq3 = (m_{\tst_1}+m_{\tst_2}+m_{\tb_1})/3
\ee
 while the $x$-axis lists the 
particular parameter. Blue points have $\msq3 <1.5$ TeV, green points have $\msq3 <1$ TeV
and red points have $\msq3 <0.5$ TeV. 

In frame {\it a}), we see that we can generate solutions with $\msq3$
lower than $0.5$ TeV, but only for values of $m_0(1,2)\alt 18$ TeV.  For
heavier values of $m_0(1,2)$, it is well known that two-loop RGE effects
tend to push third generation squark masses into the tachyonic
range\cite{am,gutimh,esusy}, which here would correspond to color
breaking minima in the scalar potential.  On the other hand, requiring
$\msq3 <1$ (1.5) TeV allows for $m_0(1,2)$ as high as $\sim 25$ TeV--
enough to suppress FCNCs except in the case of very large flavor
violating soft terms\cite{am}.  In frame {\it b}), we plot the required
value of $m_0(3)$ to give rise to sub-TeV average squark masses: here,
values of $m_0(3)<2$ (5) TeV are required to generate solutions with
$\msq3 <0.5$ (1) TeV. Frame {\it c}) shows the value of $m_{1/2}$
required for natural SUSY models. A value of $m_{1/2}<1.4$ TeV is
required for $\msq3 <0.5$~TeV, while $m_{1.2}\alt 1.7$~TeV because
we impose $m_{\tg}\alt 4$ TeV.  In frame {\it d}), we see
that $\msq3<0.5$ TeV can only be achieved for $A_0\agt 0$, while $\msq3 <1$
TeV is allowed for $A_0>-2m_0(3)$, {\it i.e.} $A_0$ cannot be large,
negative.  In frame {\it e}), we find that $\msq3<0.5$ is allowed for
$\tan\beta < 50$, while $\msq3<1$ TeV can be achieved for any
$\tan\beta$ from $\sim$~2 - 60. Finally, frame {\it f}) shows that
solutions with $\msq3<0.5$ TeV can be found for any value of
$m_A:0.15-2$ TeV.
\FIGURE[tbh]{
\includegraphics[width=7cm,clip]{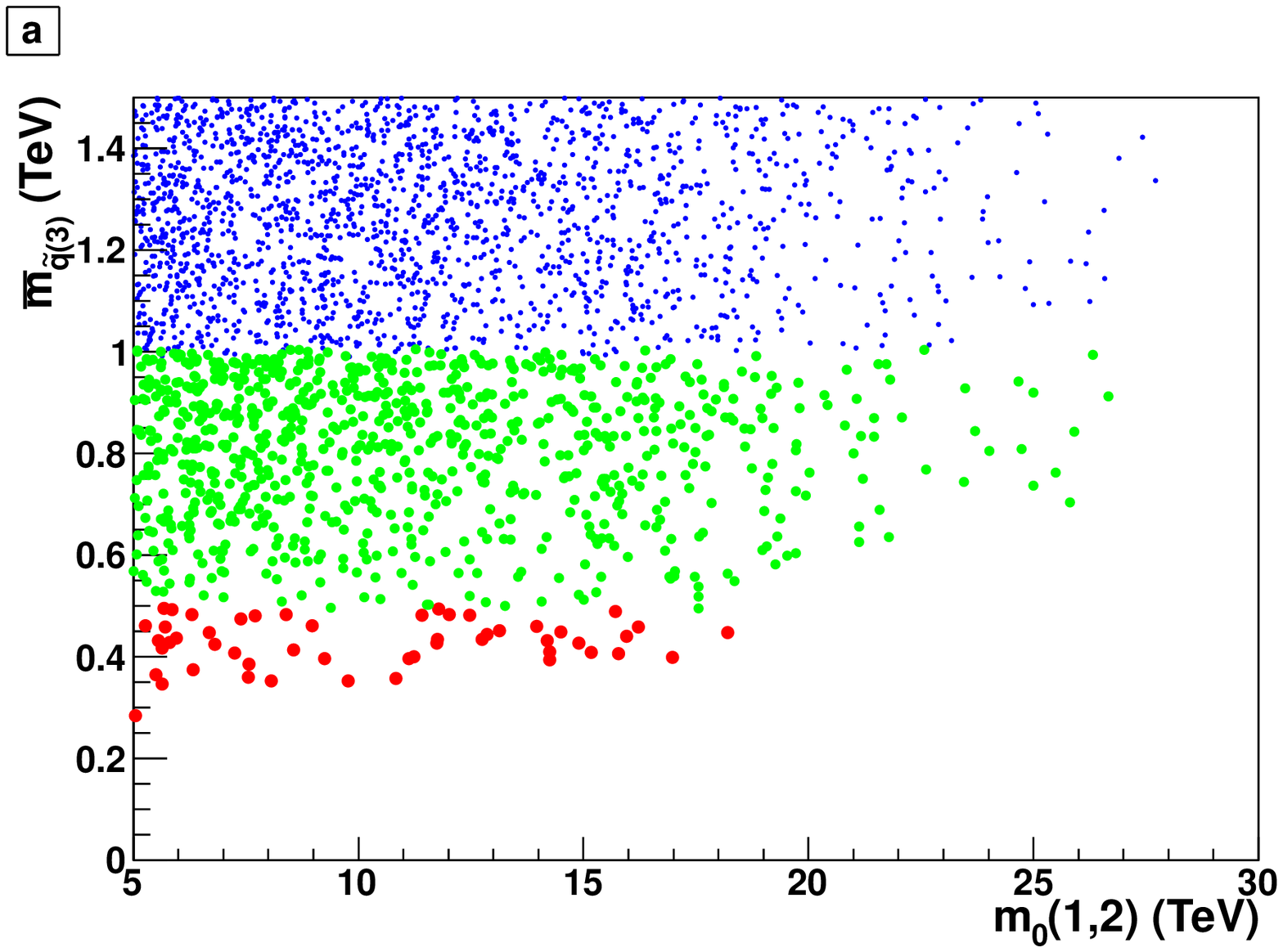}
\includegraphics[width=7cm,clip]{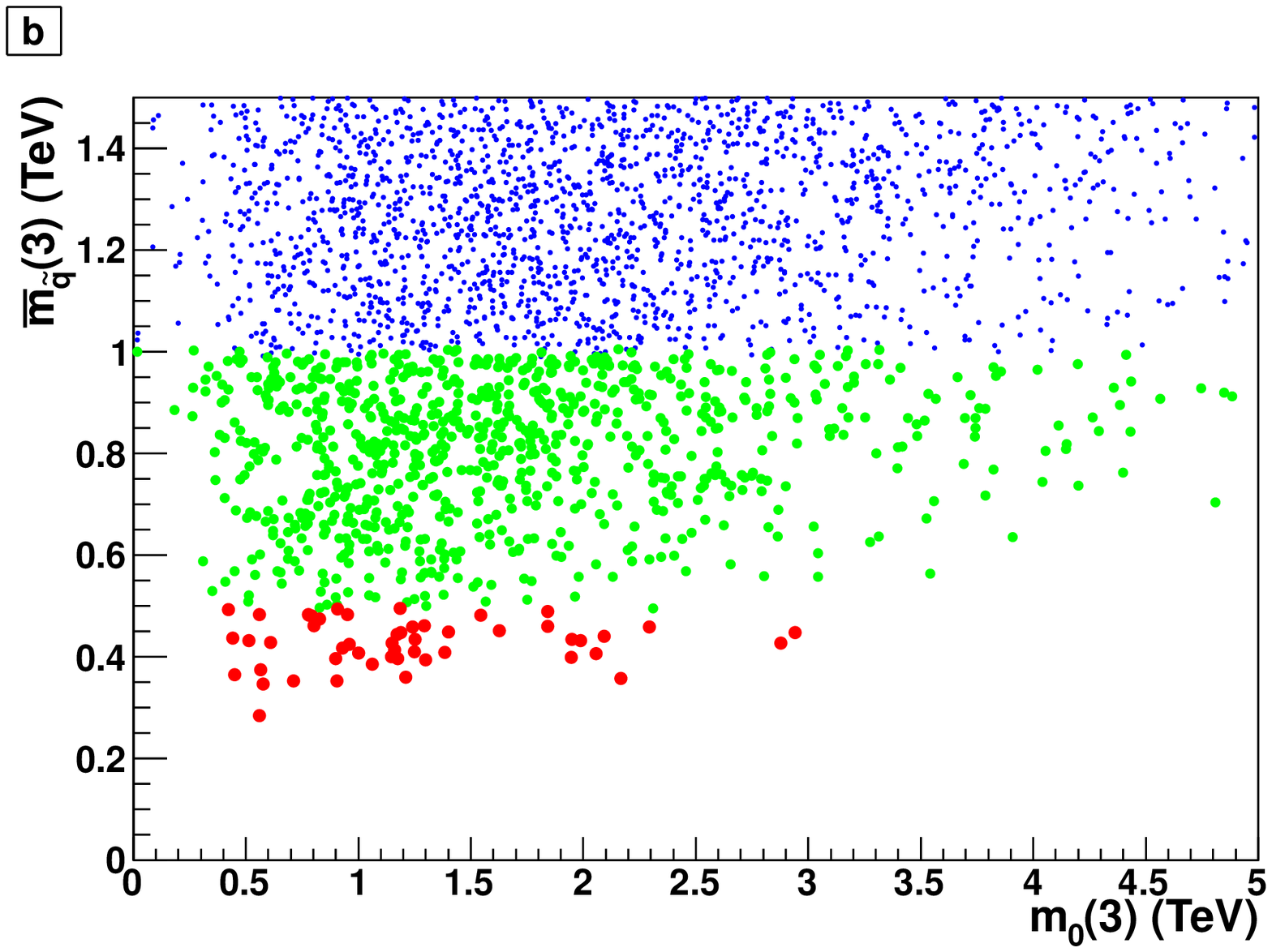}
\includegraphics[width=7cm,clip]{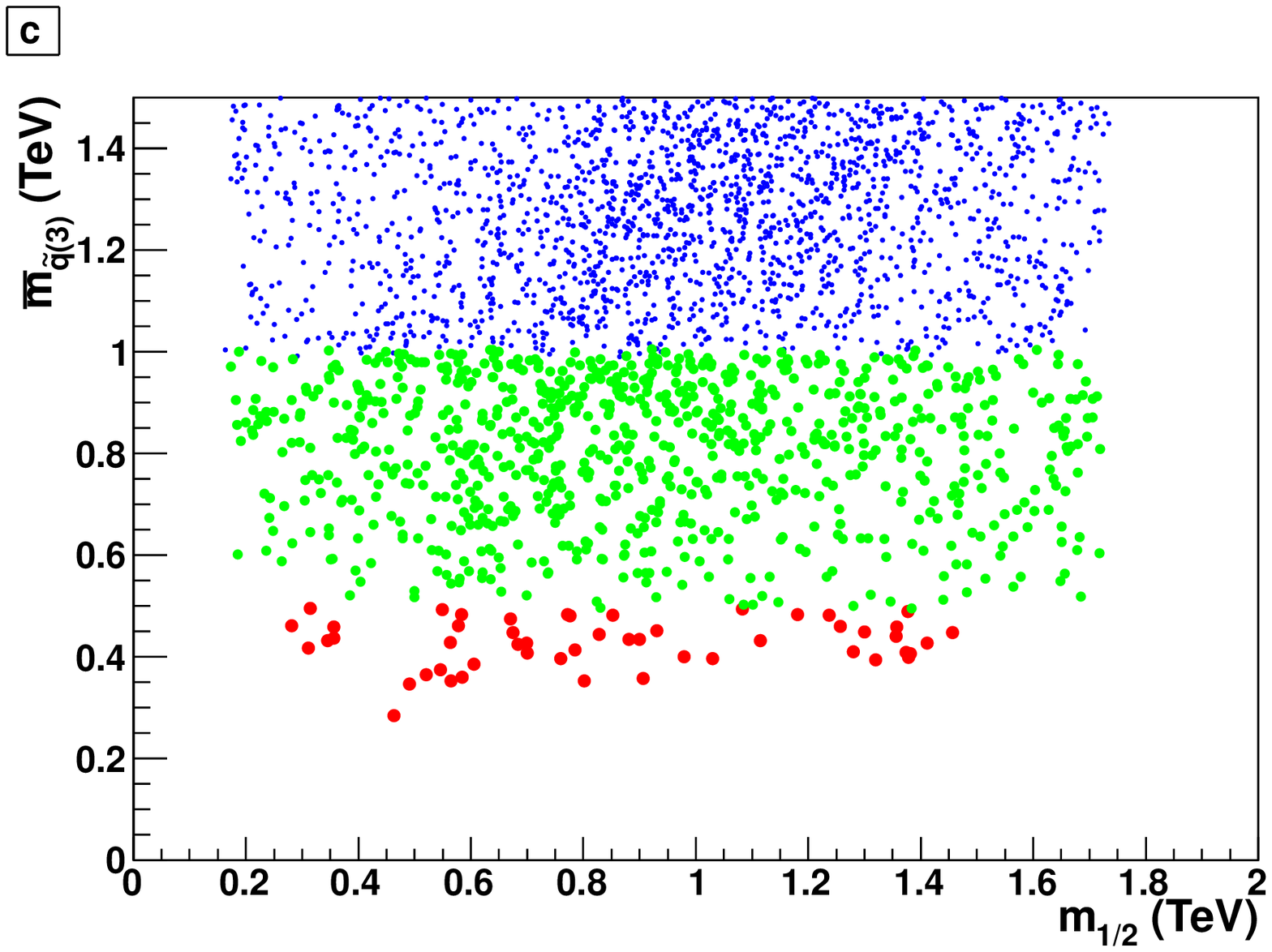}
\includegraphics[width=7cm,clip]{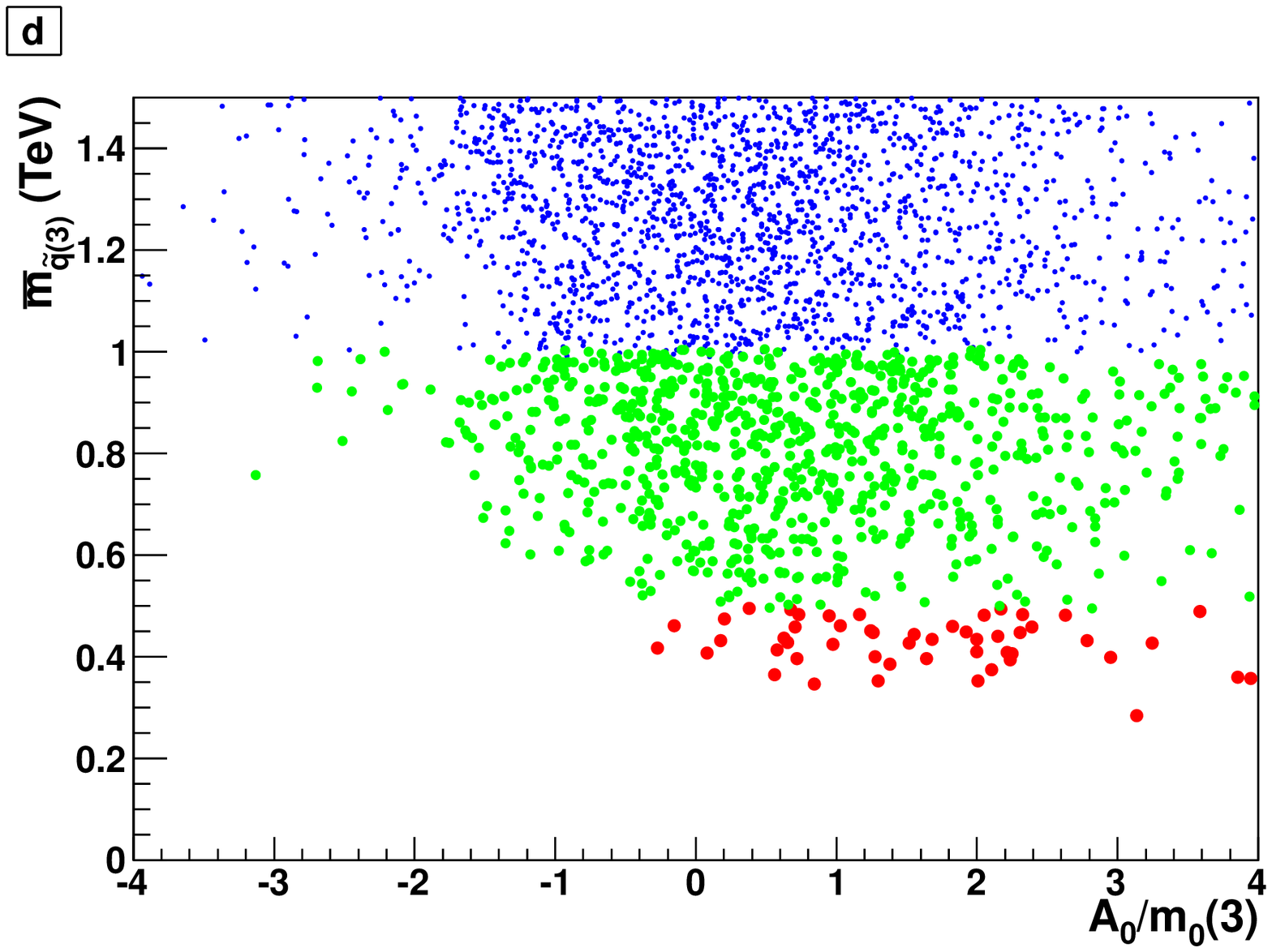}
\includegraphics[width=7cm,clip]{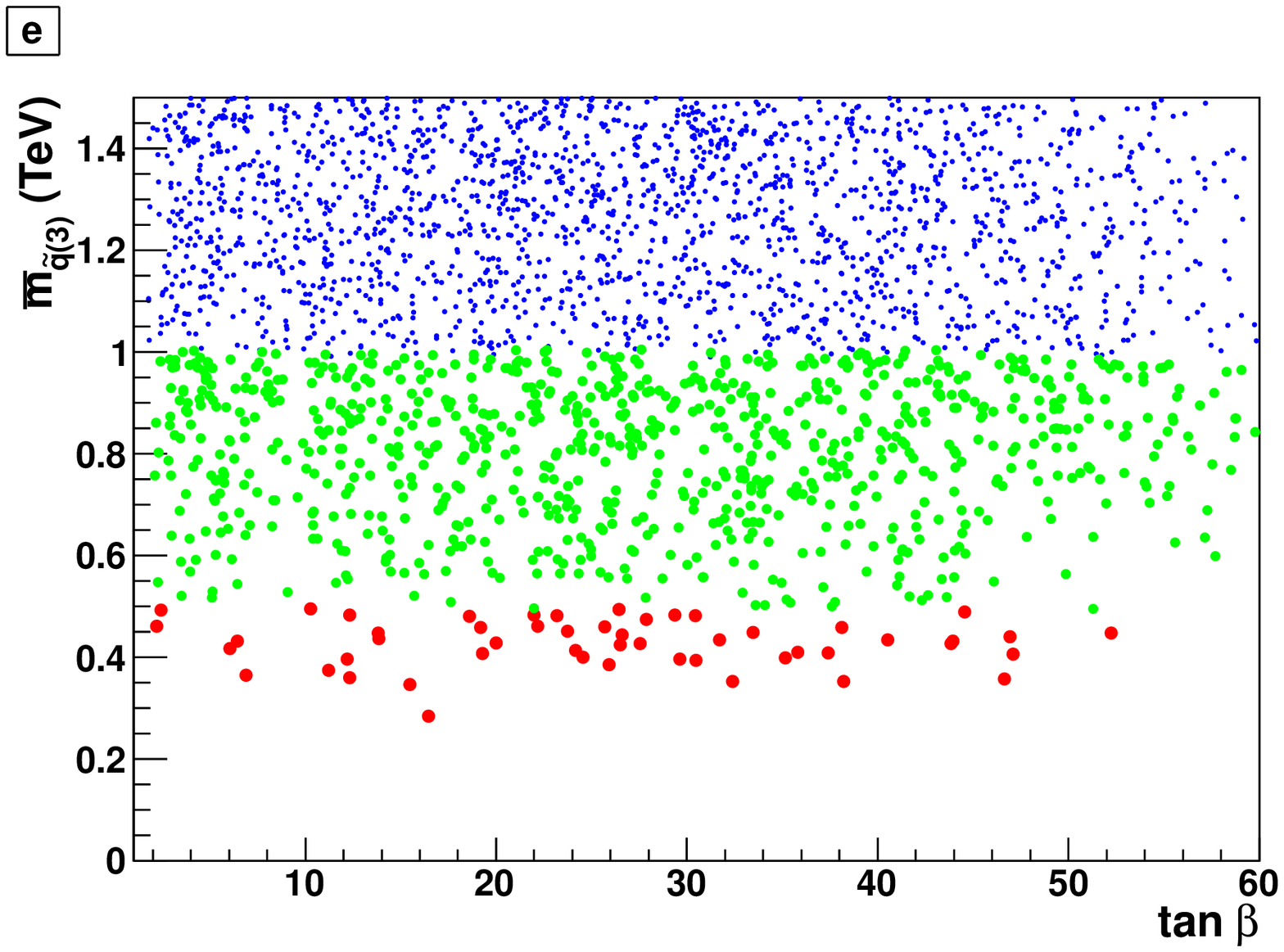}
\includegraphics[width=7cm,clip]{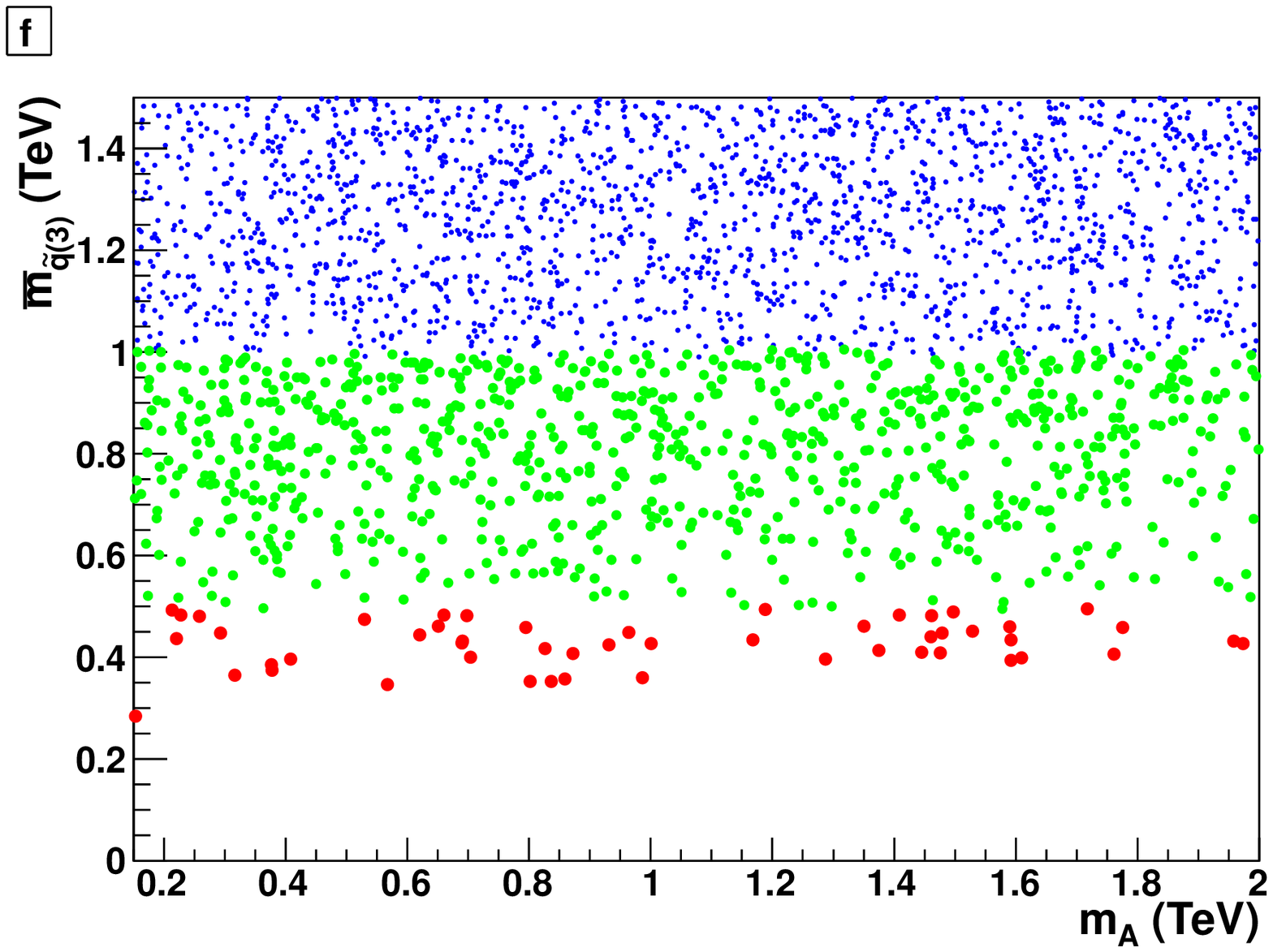}
\caption{The value of $\msq3$ versus various SUSY parameters with $\mu
=150$ GeV. The dots are colour-coded by the range of $\msq3$:$\leq
0.5$~TeV (red); 0.5-1~TeV (green); 1-1.5~TeV (blue).  }
\label{fig:param}}

In Fig. \ref{fig:mh}, we show the value of $m_h$ which is generated in
NS models versus various SUSY parameters. In frame {\it a}), we see that
the red points with $\msq3 <0.5$ TeV populate the range $m_h\sim
105-120$ GeV, while $m_h$ values, as obtained using Isajet, up to
123~GeV (124~GeV) can be readily accommodated for $\msq3$ up to 1~TeV
(1.5~TeV). This should be compared with 115.5-131~GeV (114-127~GeV), the
range of light Higgs boson masses currently allowed by the ATLAS (CMS)
data \cite{atlash,cmsh} at the 95\%CL. These experiments also report a
small excess of a signal at $m_h \sim 125$~GeV. For the smallest range
of $\msq3$ in the figure, it might appear that one would be hard pressed
to accommodate the LHC hint of a 124-126 GeV light Higgs scalar. Of
course, here one must keep in mind that Atlas/CMS may really be seeing a
Higgs scalar with mass closer to 124~GeV or that $m_t$ may be slightly
larger than 173.2~GeV as assumed in our calculation of the radiative
correction.  Combining this with a $\sim 3$ GeV error anticipated in the
Isasugra calculation of $m_h$ and it becomes apparent that values of
$m_h\sim 120-121$ GeV may be consistent with the Atlas/CMS $h(125)$ hint
even for small values of $\msq3$. The largest values of $m_h$ are
obtained for $\tan\beta \agt 10$.

We have already seen in Fig. \ref{fig:param}{\it d}) -- and also here in
Fig. \ref{fig:mh}{\it d})-- that only $A_0>0$ values lead to $\msq3<0.5$
TeV, while in Ref. \cite{h125,dm125} it is found that the largest values
of $m_h$ are found for $A_0\sim -2m_0$. As we allow increasing 
values of $\msq3$ consistent with our naturalness conditions, 
we see that values of $A_0\sim -2m_0(3)$ become allowed, and
consequently higher values of $m_h$ can be accommodated. This is the
case of maximal mixing in the top squark sector, which leads to maximal
$m_h$ values\cite{carena}.
\FIGURE[tbh]{
\includegraphics[width=7cm,clip]{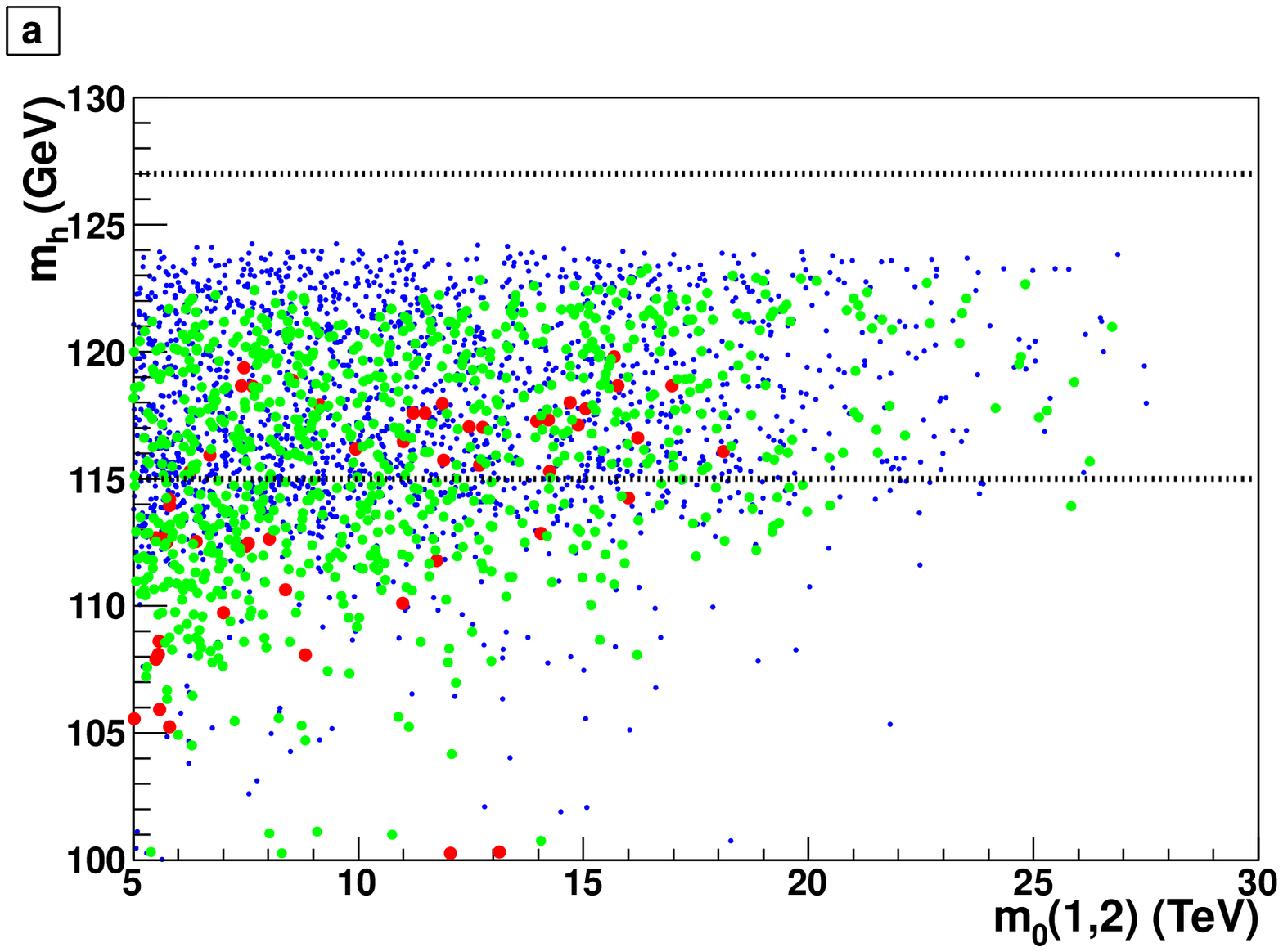}
\includegraphics[width=7cm,clip]{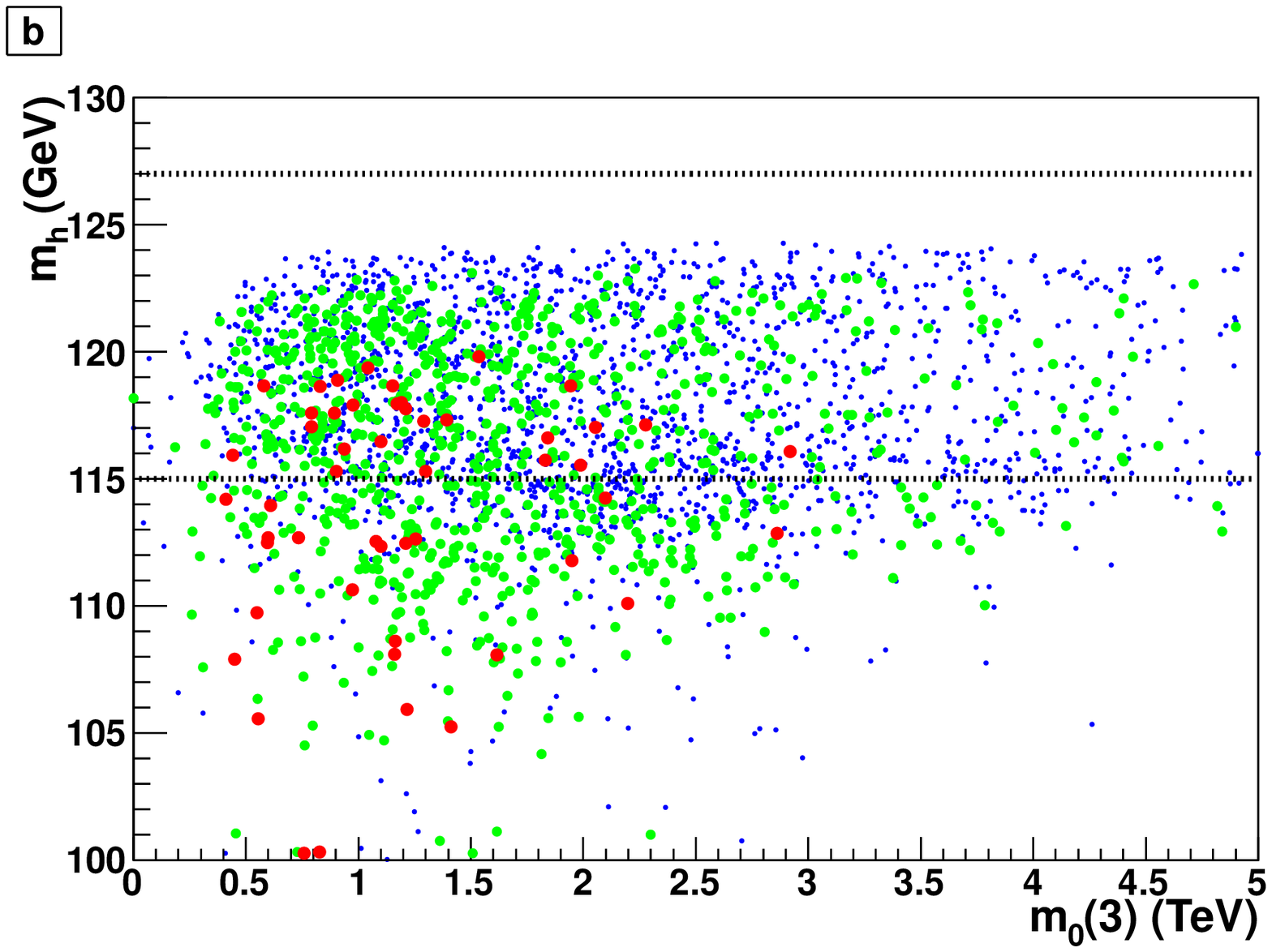}
\includegraphics[width=7cm,clip]{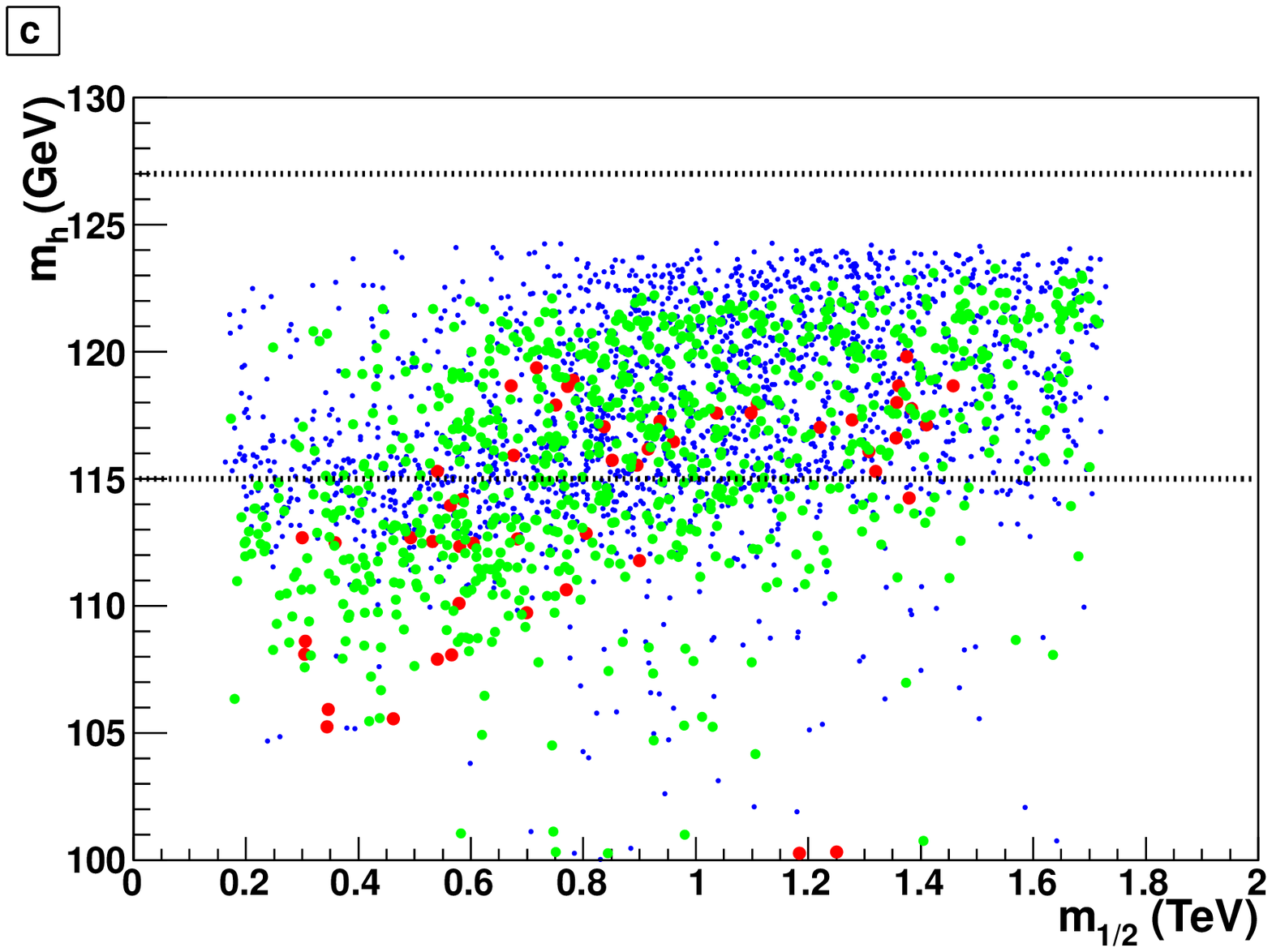}
\includegraphics[width=7cm,clip]{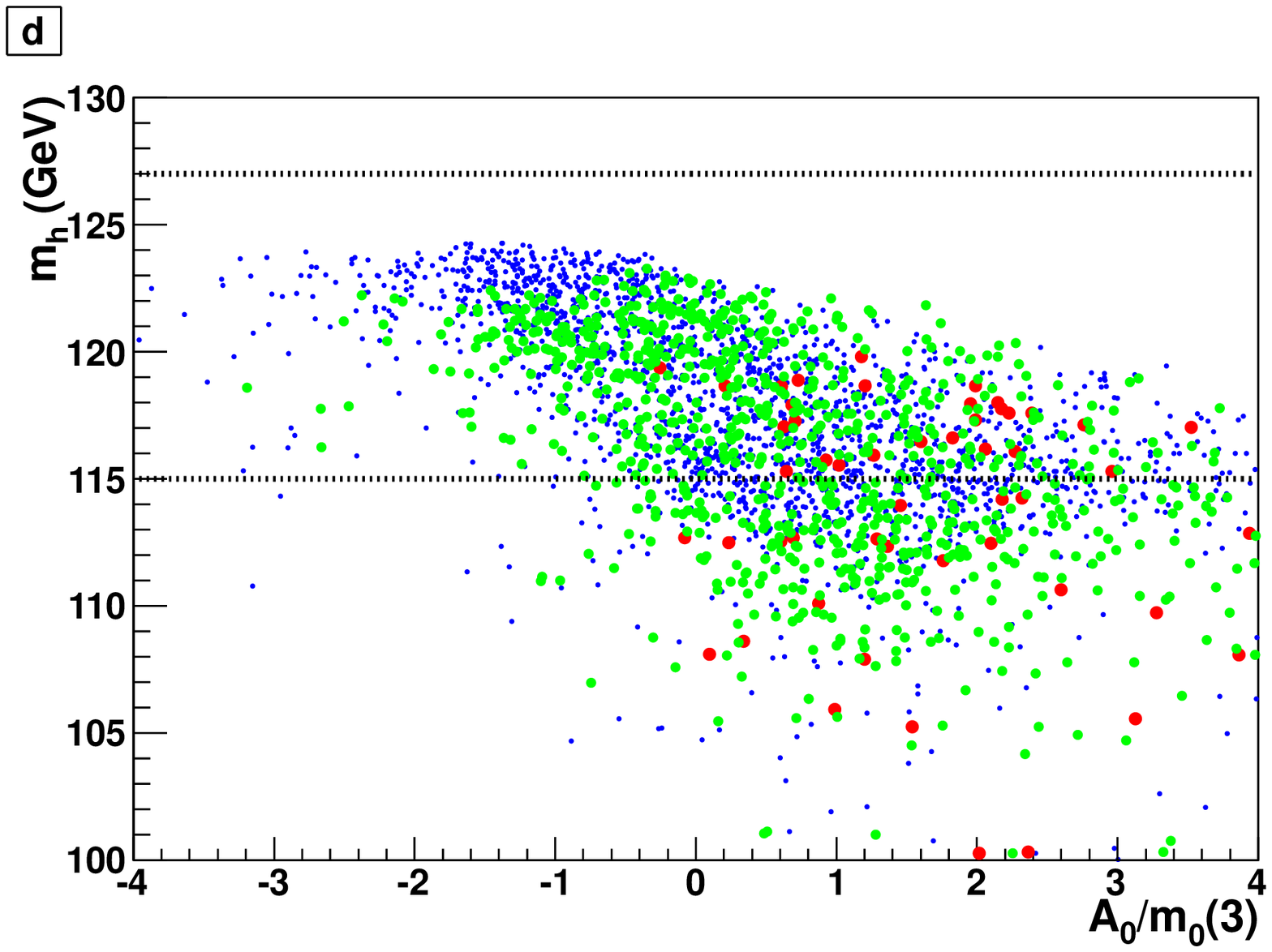}
\includegraphics[width=7cm,clip]{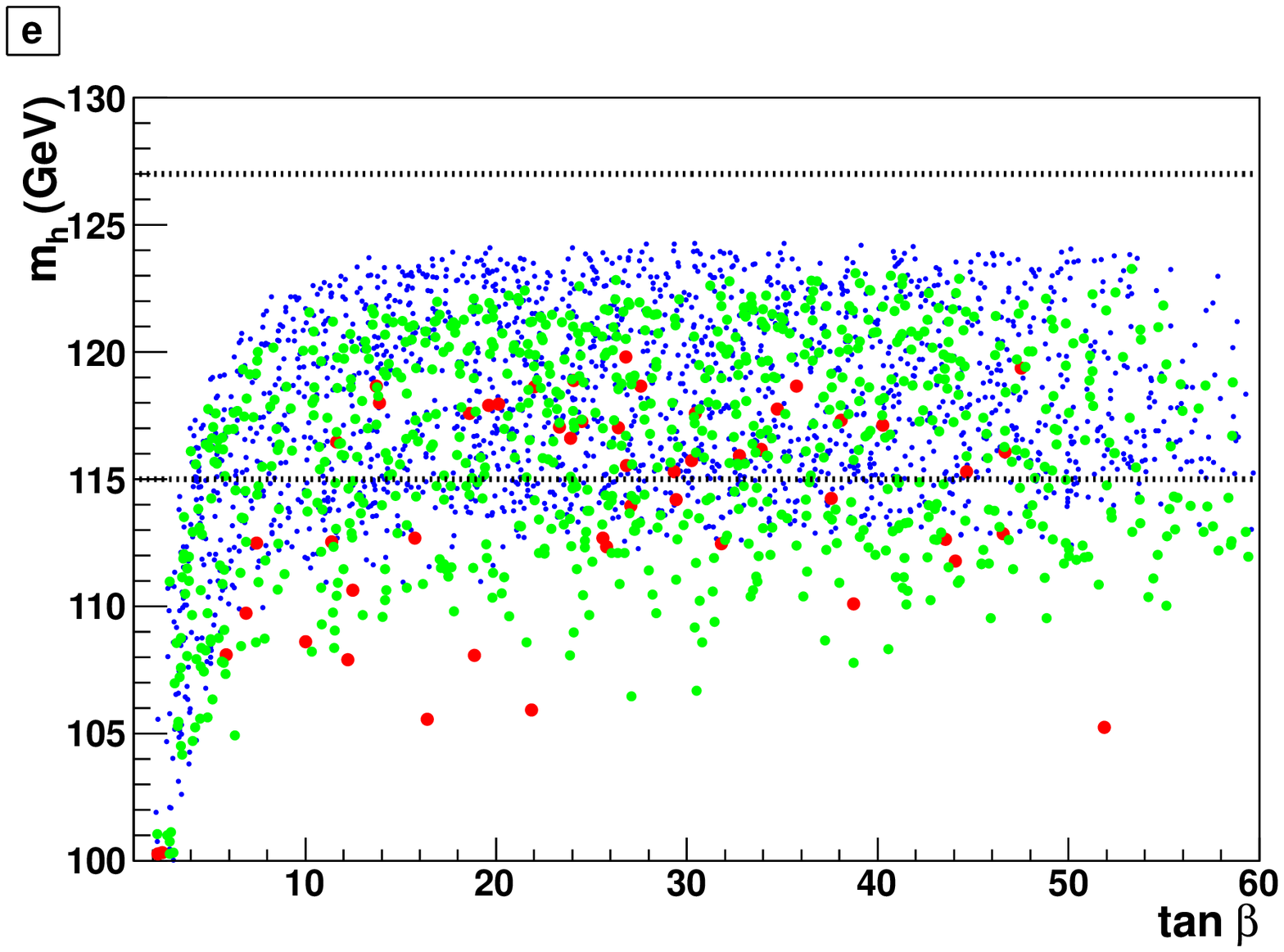}
\includegraphics[width=7cm,clip]{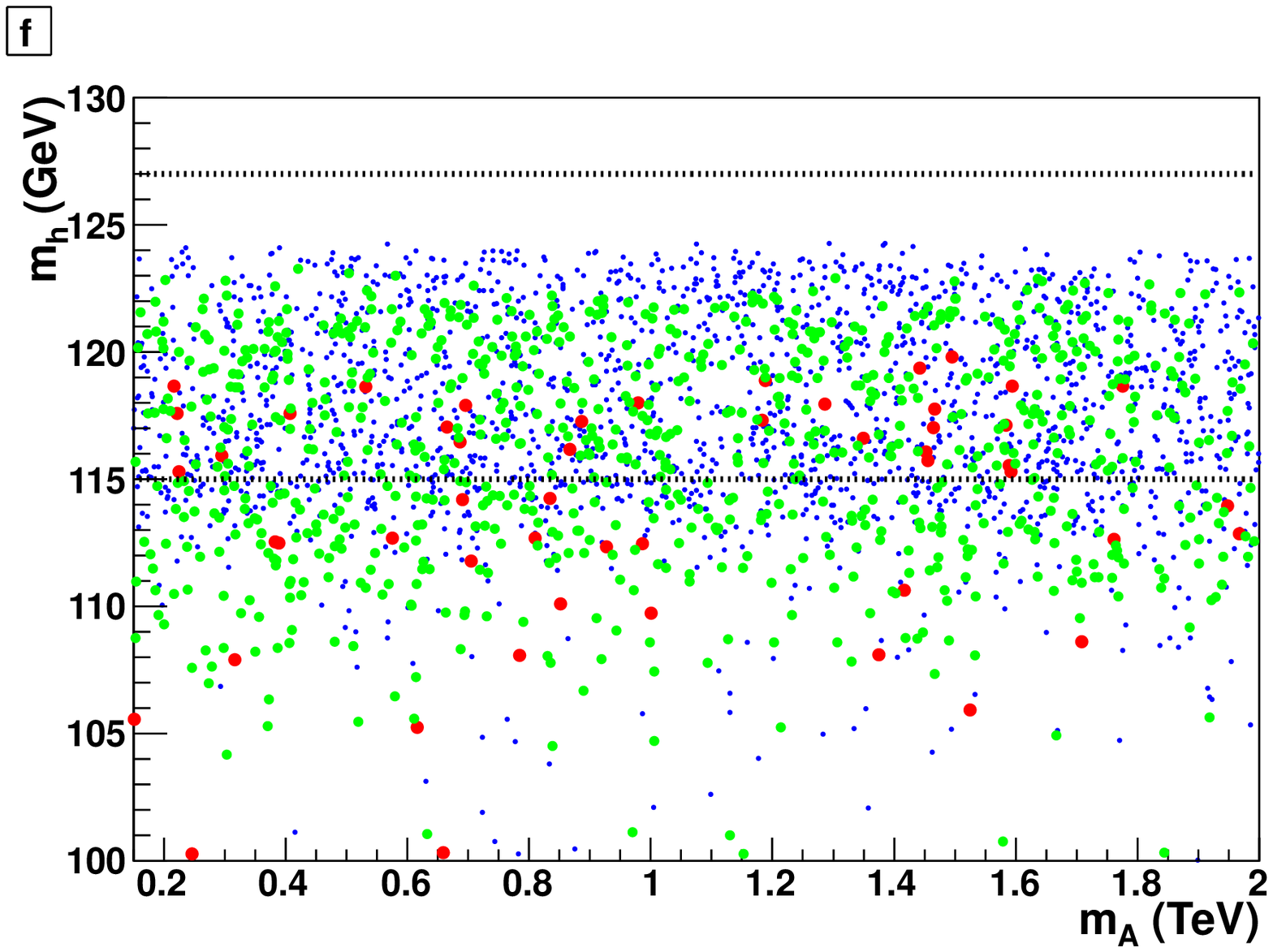}
\caption{The value of $m_h$ versus various GUT scale SUSY
  parameters. Here, and in subsequent figures, the colour-coding is as
  in Fig.~\ref{fig:param} }
\label{fig:mh}}

The value of $BF(b\to s\gamma )$ should be rather tightly constraining
for models of natural SUSY, since there may be several light third
generation squarks, and not too heavy charginos, and since the main
non-standard contributions to the decay rate come from
top-squark-chargino loops\cite{isabsg}.  Here, we implement the Isatools
subroutine IsaBSG\cite{isabsg} to compute the branching fraction, which
is listed in Figure \ref{fig:bsg} versus $\msq3$.  These values are to
be compared with the measured value of $BF(b\to s\gamma )=(3.55\pm
0.26)\times 10^{-4}$ from Ref. \cite{bsg_ex}.  Indeed, as can be seen,
large SUSY loop contributions cause the branching fraction to vary over
a wide range: $(0-9)\times 10^{-4}$, so that many solutions would be
rejected. Nonetheless, many other solutions do remain within the $\pm
3\sigma$ band (which is shown), where the various loop contributions may
cancel one against another to yield consistency with the measured value.
\FIGURE[tbh]{
\includegraphics[width=13cm,clip]{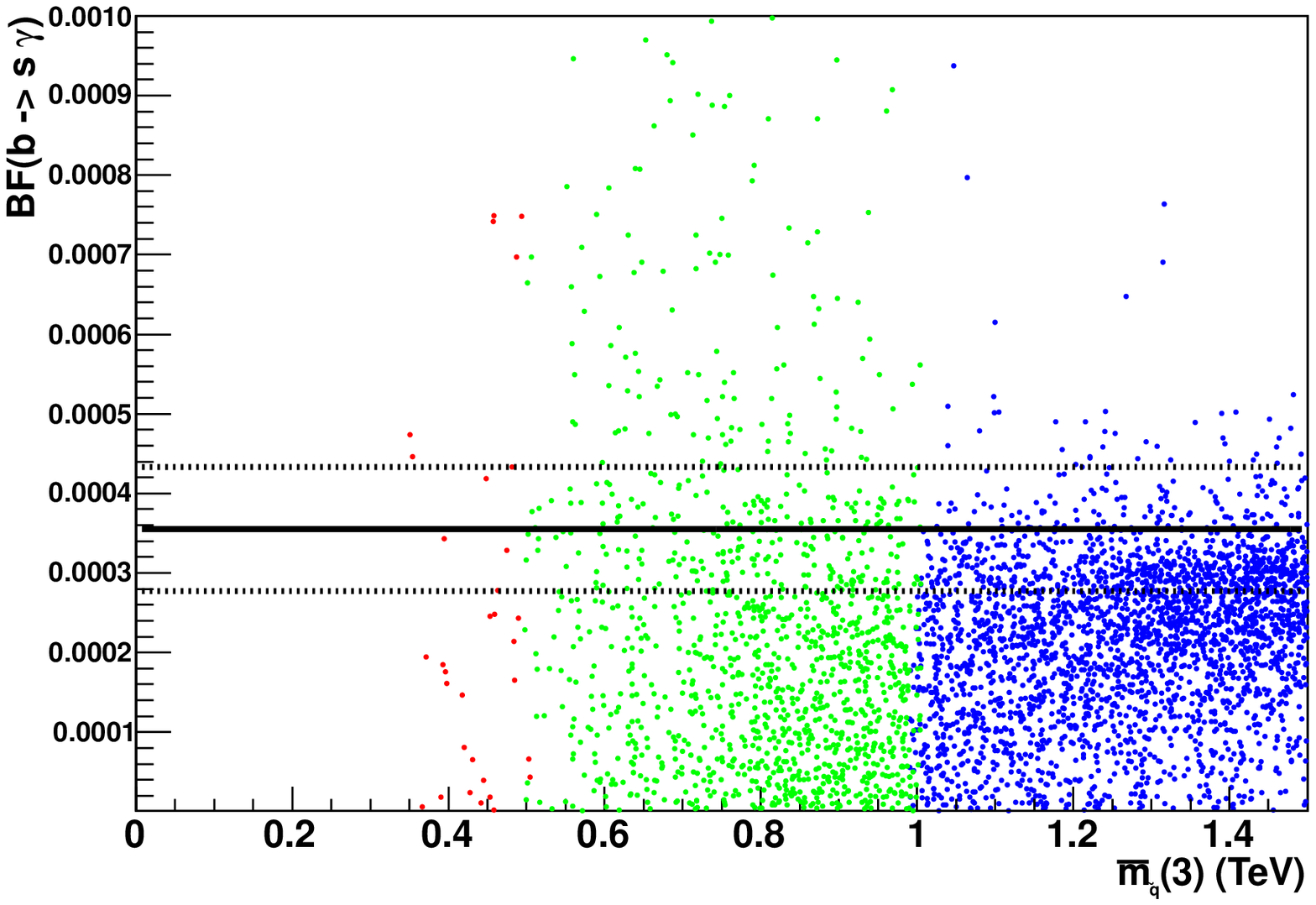}
\caption{Predicted values of the branching fraction for $b\to s\gamma$
vs. $\msq3$.  We also show the experimentally determined central
value $\pm 3\sigma$ band for the $BF(b\to s\gamma )$.  }
\label{fig:bsg}}

In addition, the well-known $(g-2)_\mu$ anomaly has been reported as a
roughly $3\sigma$ deviation from the SM value: $\Delta a_\mu=(28.7\pm
8.0)\times 10^{-10}$\cite{gm2_ex}.  In Natural SUSY, since the
$\tmu_{1,2}$ and $\tnu_\mu$ masses are in the multi-TeV range, only a
tiny non-standard contribution to the $(g-2)_\mu$ anomaly is expected,
and alternative explanations for this anomaly would have to be sought.

\section{Benchmark points, slopes and planes}
\label{sec:bm}

In this section, we list some representative natural SUSY benchmark
points, slopes and planes which could be used for LHC analyses.  In
Table \ref{tab:bm}, we show three such points, NS1 with $\msq3 =666$
GeV, NS2 with $\msq3 =595$ GeV and NS3 with $\msq3 =1343.7$ GeV. For all
points, we fix $\mu =150$ GeV, with large $m_{1/2}$ so that the higgsino-like
chargino and the two lightest higgsino-like
neutralinos have masses $\sim 150$ GeV and are
the lightest sparticles. The light Higgs masses $m_h\sim 121$ GeV
for the first two points, and so are low but as discussed above not 
incompatible with the recent hint for 
$m_h\sim 125$ GeV.  
The third point NS3 allows $m_h=123.5$ GeV but at the
expense of rather large $\msq3$, and $m_{\tst_2}$ marginally beyond our
naturalness requirement. For all these points the gluino mass is around
3~TeV and first and second generation squarks are completely beyond the
reach of the LHC. 

For point NS1, the light top squark $\tst_1$ is next-lightest SUSY
particle after the three higgsino-like states; it has mass
$m_{\tst_1}=301.4$ GeV and may be accessible to LHC top squark searches.
The $\tb_1$ and $\tst_2$ come in at 788 and 909 GeV, respectively. Both
staus and the tau sneutrino are relatively light and might be accessible
at a future TeV-scale lepton-anti-lepton collider.

Point NS2 has a light bottom squark with $m_{\tb_1}=497.3$ GeV as
next-lightest after the higgsinos.  The $\tst_1$ is slightly heavier at
572 GeV. This point has heavier tau sleptons which would not be
accessible to any planned lepton colliders.  Point NS3 with rather heavy
third generation squarks and sleptons would be very challenging to see
at LHC although the spectrum of light higgsinos should be accessible to a
linear $e^+e^-$ collider.
%
\begin{table}\centering
\begin{tabular}{lccc}
\hline
parameter & NS1 & NS2 & NS3 \\
\hline
$m_0(1,2)$      & 13363.3 & 19542.2 & 7094.3 \\
$m_0(3)$      & 761.1 & 2430.6 & 890.7 \\
$m_{1/2}$  & 1380.2 & 1549.3 & 1202.6 \\
$A_0$      & -167.0 & 873.2 & -2196.2 \\
$\tan\beta$& 22.9 & 22.1 & 19.4 \\
$\mu$      & 150 & 150  & 150 \\
$m_A$      & 1545.6 & 1652.7 & 410.1 \\
\hline
$m_{\tg}$   & 3272.2 & 3696.8 & 2809.3  \\
$m_{\tu_L}$ & 13591.1 & 19736.2 &  7432.9 \\
$m_{\tu_R}$ & 13599.3 & 19762.6 & 7433.4 \\
$m_{\te_R}$ & 13366.1 & 19537.2 & 7086.9 \\
$m_{\tst_1}$& 301.4 & 572.0 &  812.5 \\
$m_{\tst_2}$& 909.2 & 715.4 &  1623.2 \\
$m_{\tb_1}$ & 788.1 & 497.3 & 1595.5 \\
$m_{\tb_2}$ & 1256.2 & 1723.8 & 1966.7 \\
$m_{\ttau_1}$ & 430.9 & 2084.7 & 652.2 \\
$m_{\ttau_2}$ & 532.9 & 2189.1 & 1065.5 \\
$m_{\tnu_{\tau}}$ & 402.3 & 2061.8 & 1052.1 \\
$m_{\tw_2}$ & 1180.2  & 1341.2 & 1013.9 \\
$m_{\tw_1}$ & 155.9  & 156.1  & 156.2\\
$m_{\tz_4}$ & 1181.3 & 1340.4  & 1020.0 \\ 
$m_{\tz_3}$ & 615.3 & 698.8  & 532.6 \\ 
$m_{\tz_2}$ & 156.3 & 156.2 & 157.0 \\ 
$m_{\tz_1}$ & 148.4 & 149.2 & 147.4 \\ 
$m_h$       & 121.3 & 121.1 &  123.5 \\ 
\hline
$\Omega_{\tz_1}^{std}h^2$ & 0.007 & 0.006 & 0.007 \\
$BF(b\to s\gamma)$ & $2.8\times 10^{-4}$  & $3.6\times 10^{-4}$ & $2.8\times 10^{-4}$ \\
$\sigma^{SI}(\tz_1 p)$ (pb) & $5.5\times 10^{-9}$  & $1.8\times 10^{-9}$ & $9.8\times 10^{-9}$\\
$\sigma^{SD}(\tz_1 p)$ (pb) & $3.9\times 10^{-5}$  & $2.9\times 10^{-5}$ & $5.7\times 10^{-5}$\\
$\langle\sigma v\rangle |_{v\to 0}$  (cm$^3$/sec) 
& $3.0\times 10^{-25}$  & $3.1\times 10^{-25}$ & $3.0\times 10^{-25}$\\
\hline
\end{tabular}
\caption{Input parameters and masses in~GeV units for three Natural SUSY
benchmark points, with $\mu =150$ GeV. Also shown are the values of
several non-accelerator observables.  }
\label{tab:bm}
\end{table}

In Fig. \ref{fig:slope}, we convert benchmark point NS1 into a benchmark
slope by retaining all parameters as in Table \ref{tab:bm}, except
allowing $m_0(3)$ to vary. For $m_0(3)$ much below 700 GeV, we generate
spectra with tachyonic stops. Some gaps occur in the plot where no convergent RGE
solution is found. These gaps can be filled in by increasing the number of iterations
in Isasugra RGE running beyond the default value of 25. 
In the Figure, we plot all four third generation
squark masses versus $m_0(3)$, which gives a rising spectrum for most
third generation squarks except the light top squark which reaches a
minimal mass at $m_0(3)\simeq 840$ GeV, where $m_{\tst_1}<m_{\tz_1}$
so that the $\tst_1$ is lightest MSSM particle. This point gives maximal mixing in the
top squark sector, and a minimal value for $m_{\tst_1}$.
\FIGURE[tbh]{
\includegraphics[width=13cm,clip]{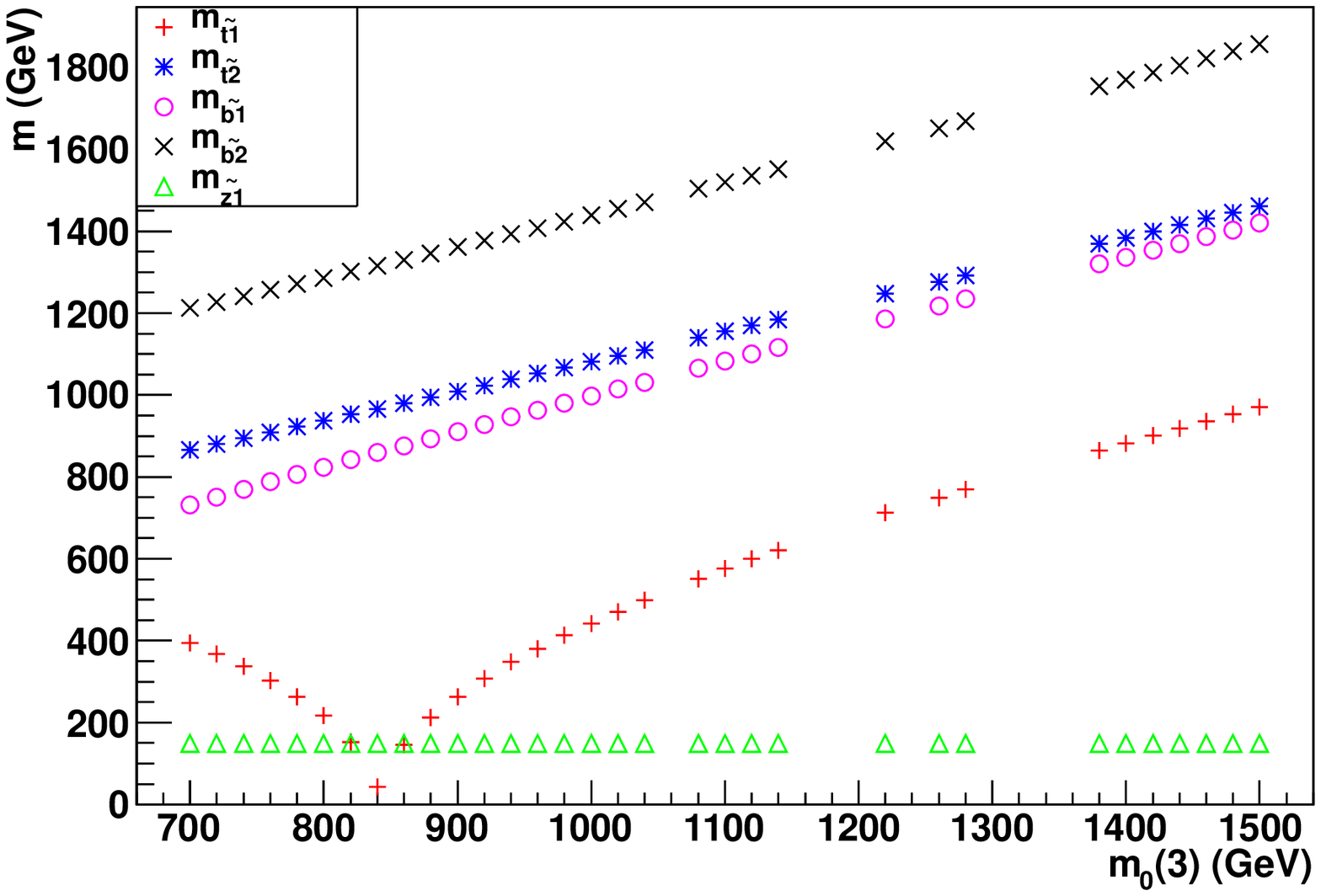}
\caption{Plot of third generation squark masses together with the lightest
  neutralino mass versus variation in
  $m_0(3)$, with other parameters fixed as for the benchmark point NS1.
}
\label{fig:slope}}

In Fig.~\ref{fig:plane}, we convert benchmark point NS1 into a benchmark
plane, where we plot contours of light top-squark mass $m_{\tst_1}$ as a
function of $m_0(3)\ vs.\ \mu$ variation. The unshaded region gives rise
to tachyonic squarks.  Points with valid solutions are labeled as black
dots; the gaps again require iterations beyond the default value of 25. 
The color coding extrapolates the generated value of
$m_{\tst_1}$, which again reaches a minimum of below $200$ GeV at
$m_0(3)\sim 830$ GeV. 
\FIGURE[tbh]{
\includegraphics[width=13cm,clip]{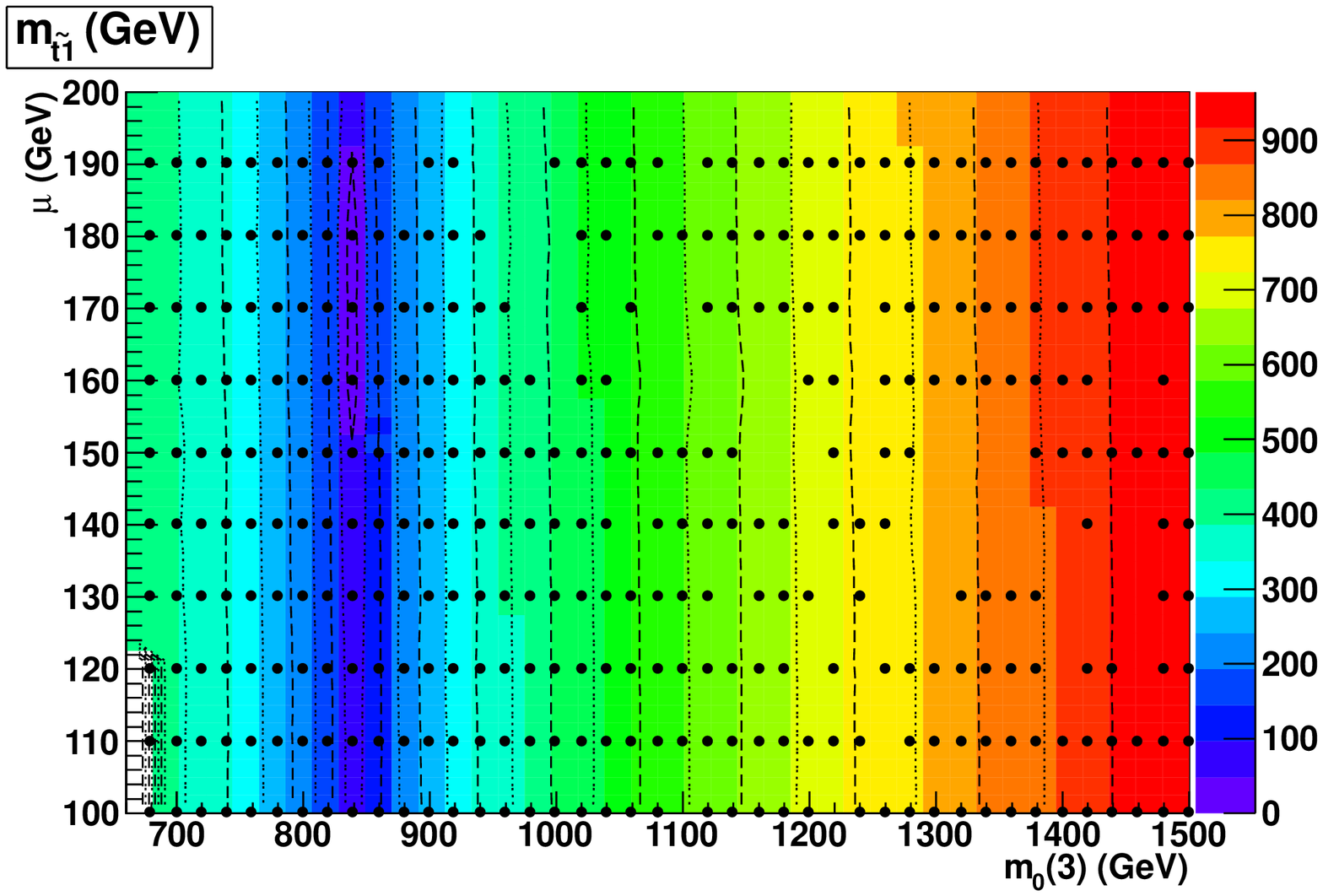}
\caption{The top squark mass $m_{\tst_1}$ in the $\mu\ vs.\ m_0(3)$ plane,
with other parameters as for benchmark point NS1.
}
\label{fig:plane}}
%

\section{LHC signals for Natural SUSY}
\label{sec:lhc}


We begin by noting  that since $\mu\alt 200$ GeV, we expect a
spectrum of light, higgsino-like $\tw_1,\ \tz_1$ and $\tz_2$ with mass
$\sim \mu$ and small mass gaps $m_{\tw_1}-m_{\tz_1}\sim
m_{\tz_2}-m_{\tz_1}\sim 10-20$ GeV.  Models with low $\mu$ parameter
have been considered previously in Refs. \cite{cheung} and
\cite{bbh}. In Ref. \cite{bbh}, production cross sections for chargino
pair production, chargino-neutralino production and neutralino pair
production were presented.  The -ino pair production cross sections tend
to be in the 50-500 fb range.  The decays $\tw_1\to\tz_1 f\bar{f}'$ and
$\tz_2\to \tz_1 f\bar{f}$ (where $f$ collectively stands for light SM
fermions) are dominated by $W^*$ and $Z^*$ exchange respectively.
However, since the mass gaps $\tw_1-\tz_1$ and $\tz_2-\tz_1$ are so
small, there is only a small visible energy release in the decays,
making the visible portion of the final state very soft and difficult to
extract above SM backgrounds. In fact, models with low $\mu$ and concomitantly
light higgsinos but other
sparticles at the 
multi-TeV scale have been dubbed ``hidden SUSY'' in
Ref. \cite{bbh} 
because distinctive SUSY signals at LHC are extremely
difficult to extract above background.

A key feature of Natural SUSY models is that they necessarily feature
three 
and possibly four relatively light 
third generation squarks. While simplified models tend to focus on the
signal from a single production mechanism, often assuming one dominant
decay channel, generally speaking in natural SUSY we expect several
third generation squarks to contribute to new physics signal
rates. Moreover, these squarks will typically have decays to all three
light higgsino-like states, and possibly also to other decay channels.
While the lightest of these squarks will have the largest production
cross sections, because of the larger mass gaps together with their more
complex cascade decays, production of heavier third generation squarks
may also yield observable signals.

In Fig. \ref{fig:prod}, we list the $pp\to\tst_1\bar{\tst}_1X$
production cross section calculated in NLO QCD using
Prospino\cite{prospino}. We show results for LHC with $\sqrt{s}=7$, 8
and 14 TeV center-of-mass energy.\footnote{More precisely, because
Prospino only allows a selection of Tevatron, LHC7 or LHC14 -- but not
LHC8 --
we have obtained the cross section for LHC8 by scaling
the corresponding Isajet cross section by the ratio of Prospino to
Isajet cross sections for LHC7.}  The $\tb_i\bar{\tb}_i$ (for $i=1,2$)
and $\tst_2\bar{\tst}_2$ cross sections are essentially identical to
those shown by making an appropriate mass substitution, since almost all
the production cross section comes from light quark $q\bar{q}$ and $gg$
fusion in the initial state. 
\FIGURE[tbh]{
\includegraphics[width=13cm,clip]{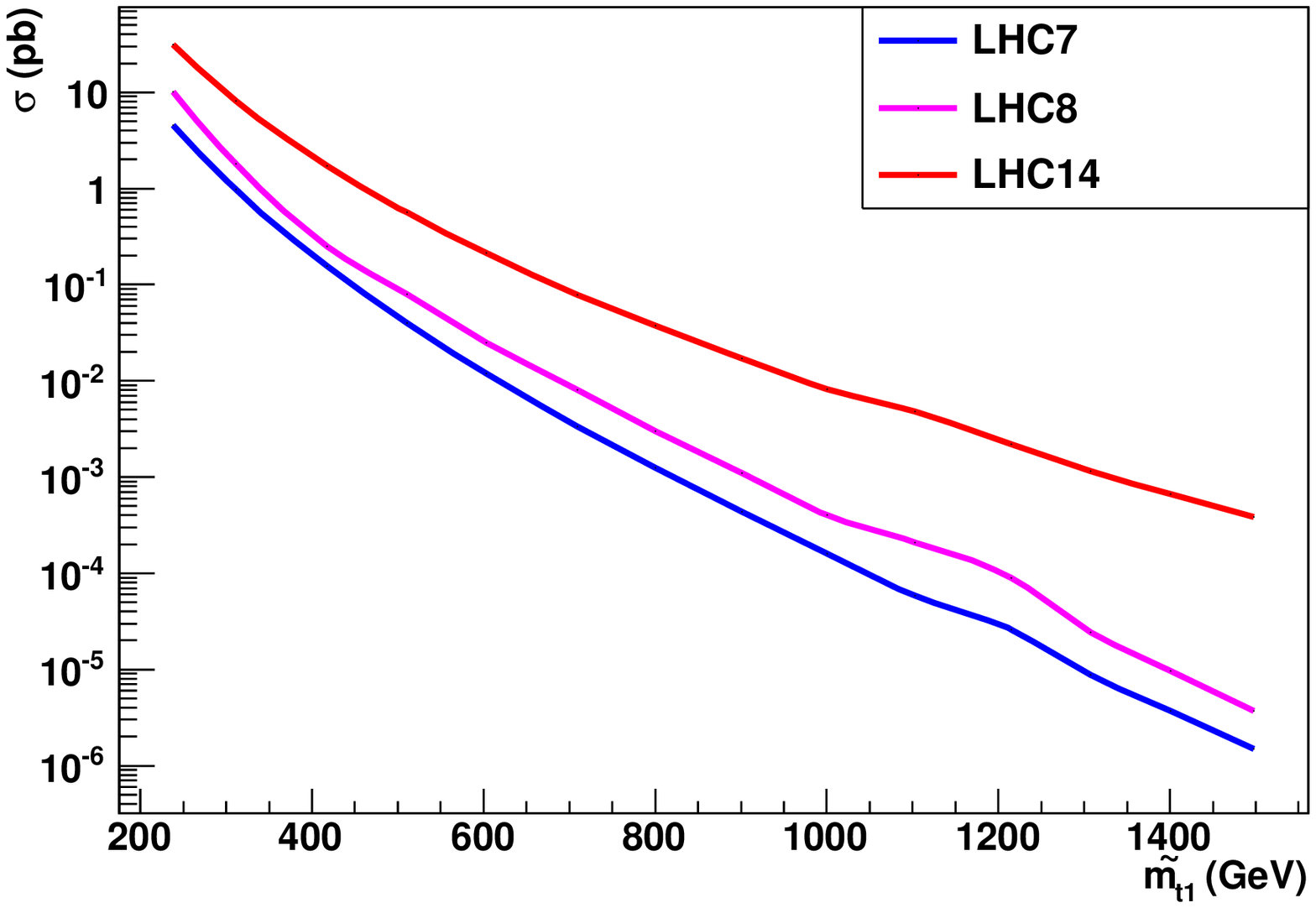}
\caption{Top squark pair production cross sections at
LHC7, LHC8 and LHC14 versus $m_{\tst_1}$. With the appropriate mass
substitution these curves also represent the cross sections for
pair production of $\tst_2, \tb_1$ or $\tb_2$ squarks.
}
\label{fig:prod}}

In Table~\ref{tab:sigbf}, we show the various third generation squark
pair production cross sections at LHC8 and branching fractions for
benchmark points NS1, NS2 and NS3 from Table \ref{tab:bm}.  Point NS1 is
by far dominated by $\tst_1\bar{\tst}_1$ production at LHC8 with a cross
section of $\sim 2$ pb. The $\tst_1$ then decays to $b\tw_1$ at $\sim 100\%$ branching fraction. 
This model would be well-described by a simplified model analysis, where the final
state $\tw_1$ is essentially regarded as missing-$E_T$ due to its soft
decay products. Thus, the signature would be a pair of acollinear
$b$-jets together with $\eslt$ and no other transverse activity except
from QCD radiation. The dominant SM physics background would be 
from $Zb\bar{b}$ production, with $Z\to\nu\bar{\nu}$. At LHC8, there is
also a 4 fb cross section from $\tb_1\bar{\tb}_1$ production followed by
$\tb_1\to W\tst_1$, giving rise to $b\bar{b}W^+W^-+\eslt$ events, albeit
at low rates.  These would be subject to a daunting background from
$t\bar{t}$ production. We mention that at LHC14, $\tst_2$-pair
production which has a cross section of $\sim 20$~fb could lead to a
handful of spectacular $\tst_2\bar{\tst}_2\to ZZ\tst_1\bar{\tst}_1 \to ZZb\bar{b}+\eslt$ 
events where the $Z$s might be identified via their leptonic decays. 
%
\begin{table}\centering
\begin{tabular}{lccc}
\hline parameter & NS1 & NS2 & NS3 \\ \hline $\sigma
(\tst_1\bar{\tst}_1)$ & $ 2000\ {\rm fb}$ & 30~{\rm fb} & 2~{\rm fb}\\
$BF(\tst_1\to b\tw_1 )$ & 1.0 & 0.25 & 0.62 \\ $BF(\tst_1\to t\tz_1 )$ &
-- & 0.42 & 0.08 \\ $BF(\tst_1\to t\tz_2 )$ & -- & 0.33 & 0.30 \\ \hline
$\sigma (\tb_1\bar{\tb}_1)$ & 4~{\rm fb} & 80~{\rm fb} & 0.00013~{\rm
fb} \\
$BF(\tb_1\to b\tz_1 )$ & 0.01 & 0.10 & 0.01 \\ $BF(\tb_1\to b\tz_2
)$ & 0.01 & 0.09 & 0.01 \\ $BF(\tb_1\to t\tw_1 )$ & 0.09 & 0.81 & 0.04
\\ $BF(\tb_1\to W\tst_1 )$ & 0.89 & -- & 0.94 \\ \hline $\sigma
(\tst_2\bar{\tst}_2)$ & 1~{\rm fb} &6~{\rm fb} & 0.00011~{\rm fb} \\ $BF(\tst_2\to b\tw_1
)$ & 0.09 & 0.29 & 0.05 \\ $BF(\tst_2\to Z\tst_1 )$ & 0.70 & 0.01 & 0.39
\\ $BF(\tst_2\to h\tst_1 )$ & 0.01 & 0.23 & 0.25 \\ $BF(\tst_2\to W\tb_1
)$ & 0.03 & 0.16 & 0.26 \\ $BF(\tst_2\to t\tz_1 )$ & 0.09 & 0.13 & 0.03
\\ $BF(\tst_2\to t\tz_2 )$ & 0.08 & 0.16 & 0.02 \\ \hline $\sigma
(\tb_2\bar{\tb}_2)$ & 0.05~{\rm fb} & 0.0001~{\rm fb} & 0.00004~{\rm fb}
\\ 
$BF(\tb_2\to
b\tz_1 )$ & 0.22 & 0.23 & 0.01 \\ $BF(\tb_2\to b\tz_2 )$ & 0.22 & 0.22 &
0.01 \\ $BF(\tb_2\to b\tz_3 )$ & 0.07 & 0.08 & -- \\ $BF(\tb_2\to t\tw_1
)$ & 0.42 & 0.44 & 0.02 \\ $BF(\tb_2\to W\tst_1 )$ & 0.03 & 0.01 & -- \\
$BF(\tb_2\to h\tb_1 )$ & 0.03 & 0.02 & -- \\ $BF(\tb_2\to H\tb_1 )$ & --
& -- & 0.23 \\ $BF(\tb_2\to A\tb_1 )$ & -- & -- & 0.23 \\ $BF(\tb_2\to
H^-\tst_1 )$ & -- & -- & 0.41 \\ $BF(\tb_2\to H^-\tst_2 )$ & -- & -- &
0.08 \\ \hline
\end{tabular}
\caption{Production cross sections at LHC8 and branching fractions
for third generation squark production
for the Natural SUSY benchmark points from Table 1.
}
\label{tab:sigbf}
\end{table}

For the point NS2, $pp\to\tb_1\bar{\tb}_1$ production is dominant at
$\sigma\sim 80$ fb, although $\tst_1\bar{\tst}_1$ is also produced at
$\sim 30$ fb. In this case, the $\tb_1$ decays dominantly via $\tb_1\to
t\tw_1$ giving rise to a $t\bar{t}+\eslt$ signature at LHC.  The decays
$\tb_1\to b\tz_1$ and $\tb_1\to b\tz_2$ also occur at $\sim 10\%$
level. The $\tst_1$ decay modes are spread somewhat evenly between
$b\tw_1$, $t\tz_1$ and $t\tz_2$ final states. Again, the small mass gap
between the $\tw_1/\tz_2$ and the LSP implies that the chargino and the
neutralino daughters are essentially invisible.  By combining all modes,
the most lucrative signature channels consist of $b\bar{b}+\eslt$ and
$t\bar{t}+\eslt$ events. The heavier $\tst_2$ decay modes are spread
among many more possibilities, including decays to $W$ and $h$ bosons in
the final state; a handful of novel events may be obtained at LHC8, but
more likely at LHC14. The $\tb_2$ state appears likely undetectable even at LHC14. 

For the benchmark point NS3, the detection of third generation squarks at LHC8 appears 
to be very difficult on account of the very low cross sections. 
Even at LHC14, the cross section for $\tst_1\tst_1$ production is just 50~fb, 
and the fact that the chargino and neutralino daughters are (nearly) invisible will make identification
of the acollinear $t\bar{t}$ and $b\bar{b}$ events from this quite
challenging. Production of $\tst_2\bar{\tst}_2$ and $\tb_1\bar{\tb}_1$ at LHC14 occurs at 
a fraction of a fb level, though the  interesting topologies that include $Z$ and $h$
production from $\tst_2$ cacsade decays may be accessible at super-LHC luminosities.

We would also like to assess the prospects for discovering the gluino
of the natural SUSY framework at the LHC.  A plot of $m_{\tg}\ vs.\
\msq3$ is shown in Fig. \ref{fig:mgl} for scan points fulfilling the
$BF(b\to s\gamma)$ constraint and also $m_h>115$ GeV. From the plot, we
see that while models with $m_{\tg}\alt 1$ TeV can be readily obtained
for $\msq3 \sim 1-1.5$~TeV, a tighter restriction of $\msq3<0.5$ (1 TeV)
typically limits $m_{\tg}\agt 2$~TeV (1~TeV).  In the case of models
with multi-TeV squarks, the LHC8 reach (which should be close to LHC7
reach\cite{lhc7}) for $\sim 20$ fb$^{-1}$ fb of integrated luminosity
extends to $m_{\tg}\sim 1$ TeV. The LHC14 reach\cite{lhc14} for 100
fb$^{-1}$ extends to $\sim m_{\tg}\sim 1.8$ TeV.  These studies have
been done within the mSUGRA model, and for LHC14 without $b$-jet tagging
which should enhance the SUSY signal in Natural SUSY models. The
increased reach in the gluino mass is projected to be up to 20\%,
depending on the details of the particle spectrum \cite{btag}.  We
conclude that while some models with large $\msq3 >1$ TeV may be
accessible to LHC gluino searches,\footnote{In this context, we note
that the ATLAS LHC7 limits \cite{marzin} from gluino-mediated stop-pair
searches do not directly apply because these rely on the same-sign
dilepton signal where the lepton may arise from either the top quark
daughter of the gluino or from the chargino daughter of the top
squark. In our case, we expect leptons from the latter source to be very
soft.} there remain many models (expecially for low $\msq3 <1$ TeV)
where gluino pair production will be beyond even the LHC14 reach.
\FIGURE[tbh]{
\includegraphics[width=13cm,clip]{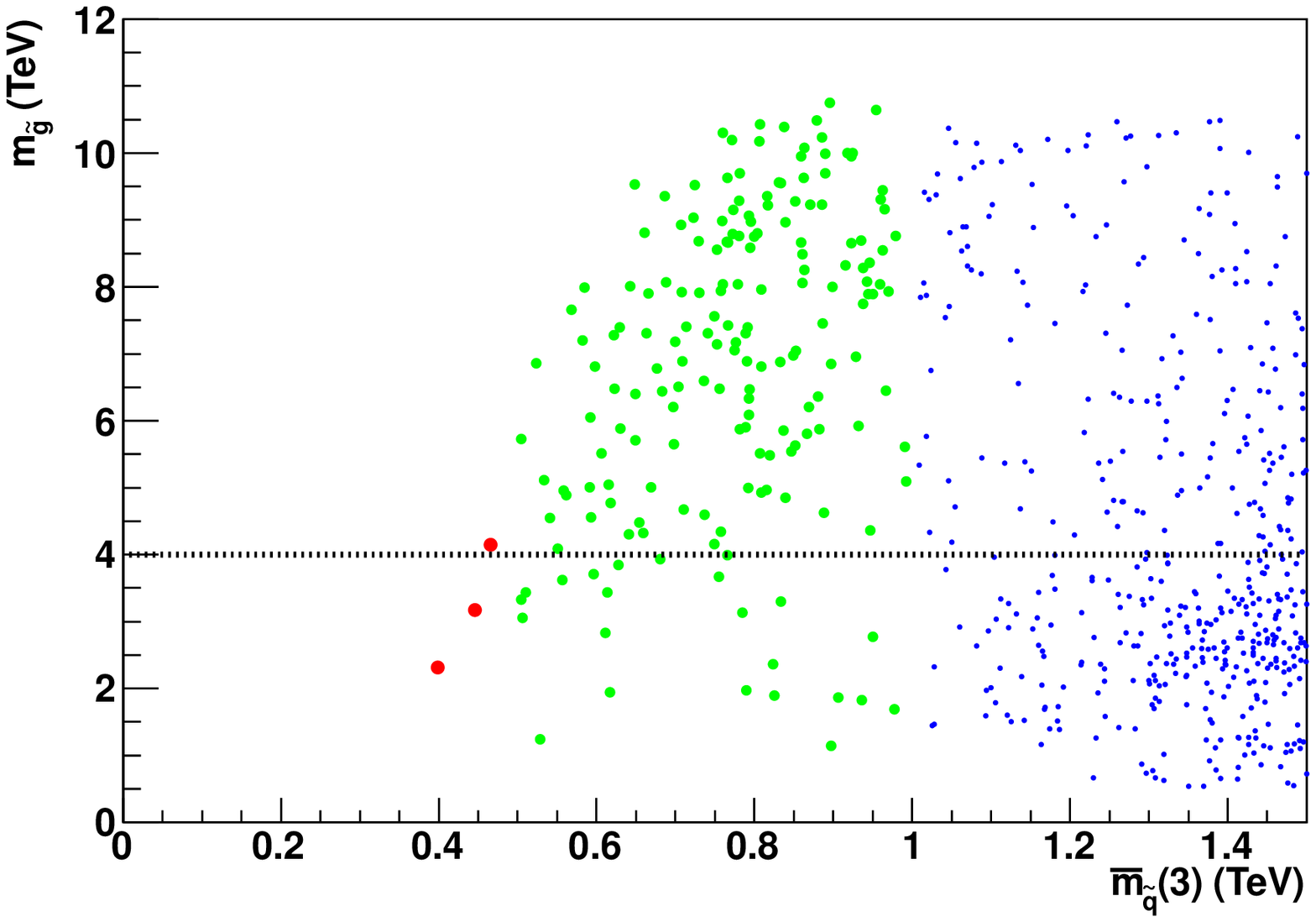}
\caption{Value of $m_{\tg}\ vs.\ \msq3$ from Natural SUSY models which obey $BF(b\to s\gamma )$ at 
$3\sigma$ and $m_h>115$ GeV.
}
\label{fig:mgl}}

Lastly, motivated by the bound on $m_A$ presented in Sec.~\ref{ssec:nat}, 
we plot $m_A \ vs \ \msq3$ in Fig.~\ref{fig:mA}
for natural SUSY points with $m_{\tg} < 4$~TeV, $m_h>115$ GeV and which satisfy the
$B(b\to s\gamma)$ constraint.
The color coding is as in Fig.~\ref{fig:mgl}. 
We see that $m_A\agt 500$ GeV for low $\msq3$ values, while
$m_A$ can be as low as a few hundred GeV for very large $\msq3$. 
\FIGURE[tbh]{
\includegraphics[width=13cm,clip]{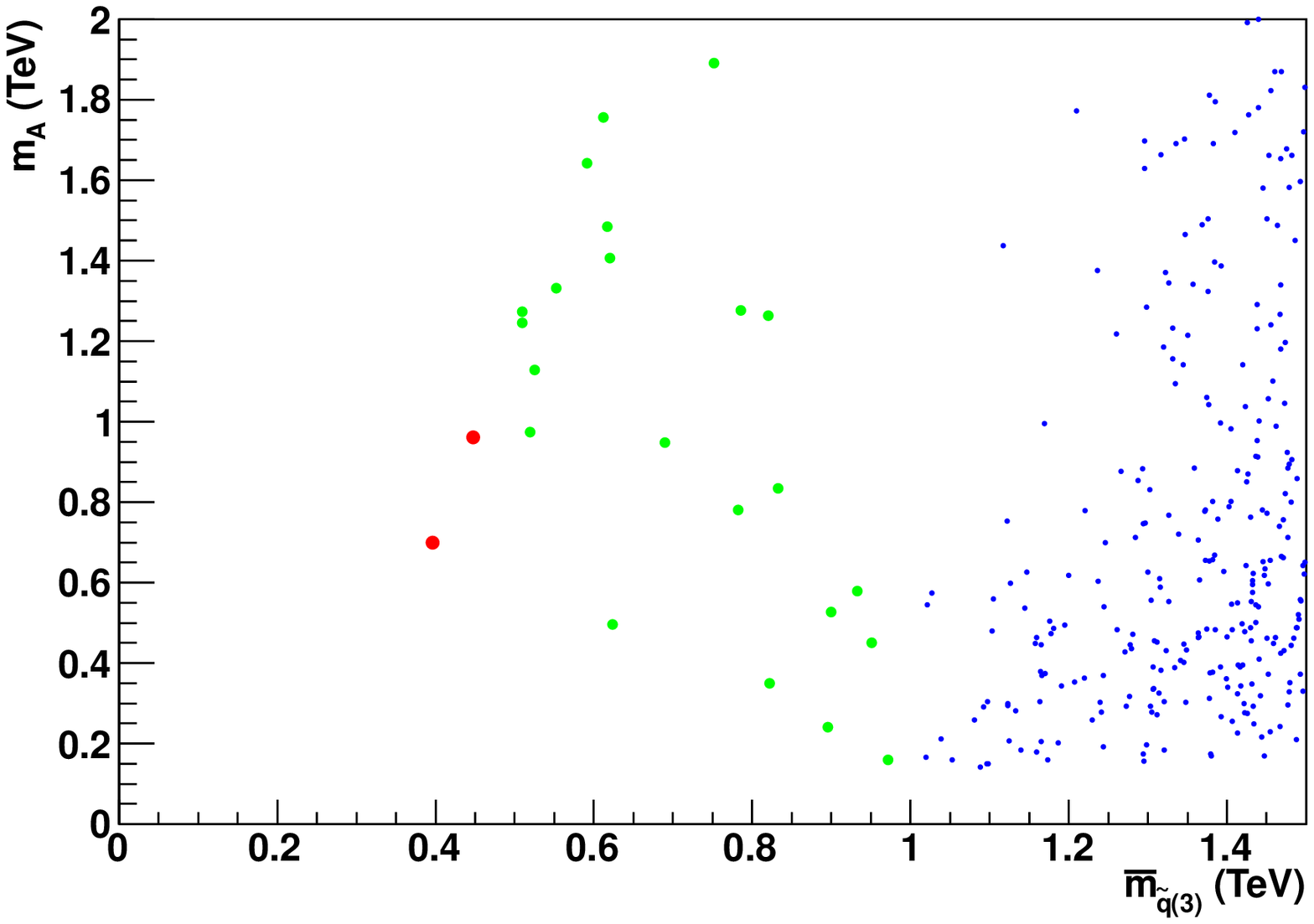}
\caption{Value of $m_{A}\ vs.\ \msq3$ in Natural SUSY models with
  $m_{\tg} < 4$~TeV, $m_h>115$ GeV and which satisfy 
$BF(b\to s\gamma )$ at $3\sigma$. The color coding is as in Fig.~\ref{fig:mgl}. 
}
\label{fig:mA}}
%

\section{Natural SUSY at a linear $e^+e^-$ collider}
\label{sec:ilc}

While Natural SUSY may possibly be difficult to discover at the LHC,
it leads to a potential bonanza of signals for 
a linear $e^+e^-$ collider (LC) operating in the $0.3-1$ TeV range. A LC would
potentially be a {\em higgsino factory} because, as we have emphasized,
the $\mu$ parameter is necessarily small in this scenario.
Indeed, pair production of higgsino-like chargino and also neutralino
states $\tw_1^\pm$, $\tz_1$ and $\tz_2$ with sizeable cross sections is
inevitable at a $0.25-0.5$~TeV machine. Example cross section  plots have been
shown in Ref.~\cite{bbh} and so will not be repeated here.
Thus, a natural target for a LC would be the pair production reactions
$e^+e^-\to\tw_1^+\tw_1^-$, $\tz_1\tz_2$ and $\tz_2\tz_2$. While the
small $\tw_1-\tz_1$ and $\tz_2-\tz_1$ mass gaps are a formidable challenge at LHC
(and may also be problematic at a LC), it has been shown\cite{tadaslc} that with specialized cuts, 
it should be possible to extract a signal above SM background at a LC. 
The visible energy from
these reactions would be low just above threshold, but as $\sqrt{s}$
increases, the decay products from $\tw_1$ and $\tz_2$ would be boosted
to higher energies. In addition, the beam polarization would be a strong
tool not only for distinguishing the signal from $W^+W^-$ backgrounds,
but  also for distinguishing between wino-like versus higgsino-like charginos\cite{bbh}.

In the case of the NS1 benchmark, after the light higgsinos are well
studied and the CM energy $\sqrt{s}$ is increased, the next target
threshold would be $\tst_1\bar{\tst}_1$ production at $\sqrt{s}\sim
2m_{\tst_1}\sim 610$ GeV. This would be followed by the tau-snuetrino
pair production threshold at $\sqrt{s}\sim 810$ GeV, with
$\tnu_\tau\to\tw_1^+\ttau_1^-$ decay. At a little higher energy,
$\tau_1^+\ttau_1^-$ pair production would turn on, followed mainly by
$\ttau_1\to \tz_1\tau$ and $\tz_2\tau$ decay.  For the heavier spectra
shown in NS2 and NS3, the light higgsino pair production reactions would
still be available, but CM energies of over 1 TeV would be required to
pick up any squark pair production reactions. As emphasized above, the
accessibility of higgsino-like states is a generic feature of Natural
SUSY models.

\section{Natural SUSY and direct/indirect WIMP searches}
\label{sec:dm}

As noted in Sec. \ref{sec:intro}, the higgsino-like neutralinos with
masses $\sim 100-200$~GeV expected in NS models annihilate very rapidly
in the early universe and so yield a thermal relic underabundance of
CDM.  However, the neutralino relic abundance can be boosted to match
its observed value in models where

\bi
\item a PQ solution to the strong CP problem is invoked, and
thermally-produced but late-decaying axino (and/or saxion) decays
augment the SUSY particle production\cite{blrs,bls}, or
\item there exist late-decaying TeV scale moduli fields with large
branching fractions to SUSY particles that subsequently decay to the
neutralino\cite{mr}. 
\ei 

We stress that it in the first case it is not necessary that neutralinos
saturate the observed relic density. 
Indeed it is easy to select PQ parameters where the converse is true:
$\Omega_{\tz_1}h^2$ stays low while the bulk of dark matter is comprised
of axions, or even where both axion and neutralino abundances are comparable. In this
case, the direct and indirect neutralino reach estimates presented below 
(these have been obtained assuming that neutralinos saturate the CDM density)
would have to be increased by a factor of $0.1123/ \Omega_{\tz_1}h^2$. 
It is difficult, but not impossible\cite{bls}, to
lower the neutralino abundance below its standard thermally produced
prediction. Thus, we expect the above reach scale factor to be typically
between 1 and $16$, since $\Omega_{\tz_1}^{\rm thermal}h^2\sim
0.007$ for a higgsino-like relic neutralino with a mass $\sim 150$~GeV.

In Fig. \ref{fig:sigSI}, we show the spin-independent $\tz_1 p$
scattering cross section in $pb$ as obtained from IsaReS\cite{bbbo}.
Here, and in the remainder of this section we assume that $\tz_1$
saturates the DM density.  We plot points versus $\msq3$, since
$m_{\tz_1}$ is fixed typically $\sim 150$ GeV due to our choice of $\mu
=150$ GeV. We actually find that the bulk of points inhabit the $\agt
10^{-8}$ pb range.  Comparing to the bound from Xe-100\cite{aprile}, we
find that a large fraction of these points are excluded if the
higgsino-like WIMP is essentially all the dark matter. Moreover, with
this same assumption, a large fraction of surviving points lie within
the projected reach of Xe-100/2012 run, and certainly within the reach of
Xe-1ton.
\FIGURE[tbh]{
\includegraphics[width=13cm,clip]{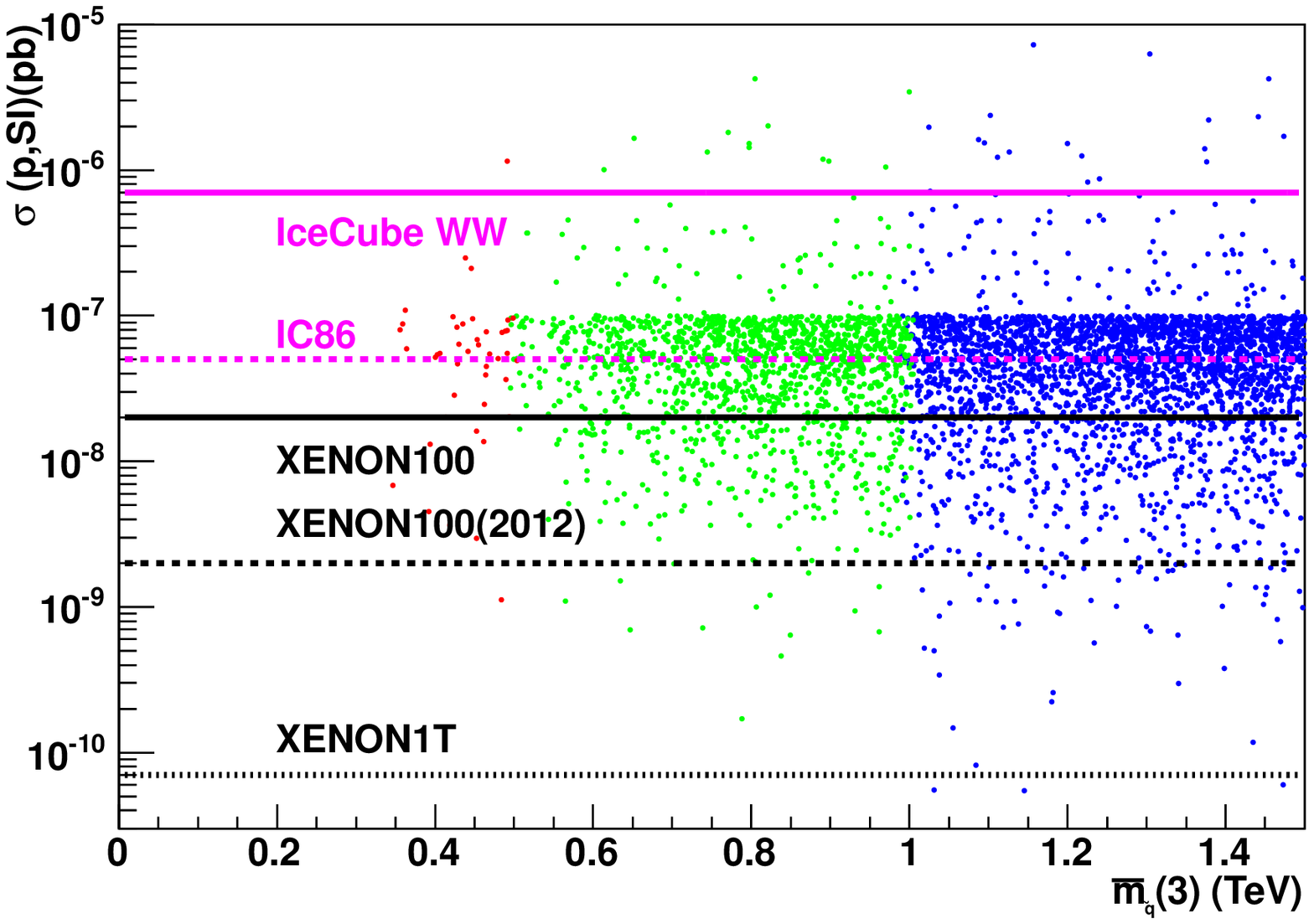}
\caption{Spin independent $p\tz_1$ scattering cross section versus
$m_0(3)$ for NS models with $\mu =150$ GeV. Also shown are the current
90\% CL bounds together with projections from the XENON100 (2012
sensitivity) and IceCube (180 day sensitivity)
experiments for a 150~GeV WIMP (assuming higgsino-like WIMPs saturate the
measured dark matter abundance).  }
\label{fig:sigSI}}

In Fig. \ref{fig:sigSD}, we plot the spin-dependent $\sigma^{SD}(\tz_1
p)$ scattering cross section, this time in comparison to current and
future IceCube reach\cite{icecube}, and future COUPP reach\cite{coupp}.
While the current IceCube reach excludes a significant portion of points
(under the asumption of neutralino dominance), the future IceCube and
especially COUPP reaches will access most of the remaining parameter
space.
\FIGURE[tbh]{
\includegraphics[width=13cm,clip]{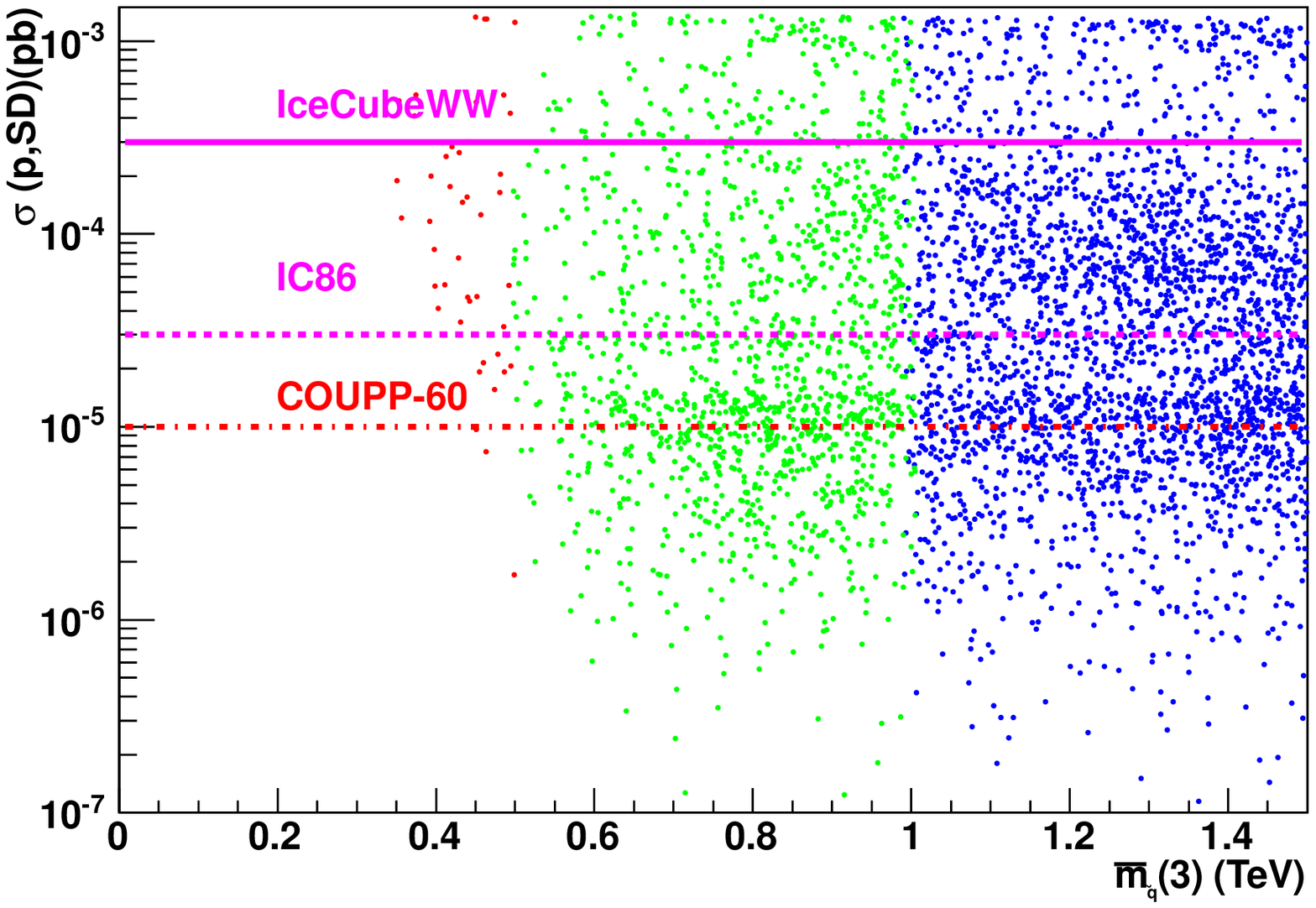}
\caption{Spin dependent $p\tz_1$ scattering cross section versus
$m_0(3)$ for NS models with $\mu =150$ GeV. Also shown are current
limits and future projections from IceCube and COUPP experiments
(assuming higgsino-like WIMPs saturate the measured dark matter
abundance).  }
\label{fig:sigSD}}

Fig. \ref{fig:sigv} shows the thermally averaged neutralino annihilation
cross section times relative velocity, evaluated as $v\to 0$. This
quantity enters linearly into indirect searches for neutralino
annihilation in the cosmos into $\gamma$s or $e^+$, $\bar{p}$ or
$\bar{D}$ searches. For our case, the bulk of points inhabit the range
$\langle\sigma v\rangle |_{v\to 0}\agt 10^{-25}$~cm$^3$/sec.  The
horizontal solid line shows the upper limit on the annihilation cross
section times velocity for very non-relativisitic dark matter in dwarf
spheroidal satellite galaxies of the Milky Way annihilating to $W$ boson
pairs obtained by the Fermi collaboration\cite{fermi}, assuming a $\sim
150$ GeV WIMP. Models with a larger annihilation cross section would
have led to a flux of gamma rays not detected by the experiment,
assuming a Navarro-Frenk-White profile for each dwarf galaxy in the
analysis.  We see that the Fermi bound excludes the bulk of points for
our choice of DM mass, again assuming higgsinos saturate the DM
density. Moreover, this bound changes rather slowly with the DM mass,
being just a factor of 2 weaker for a WIMP mass of 300~GeV.  Further
searches and improvements by the Fermi-LAT Collaboration and/or the
impending AMS results should provide an incisive probe of the NS
framework.
\FIGURE[tbh]{
\includegraphics[width=13cm,clip]{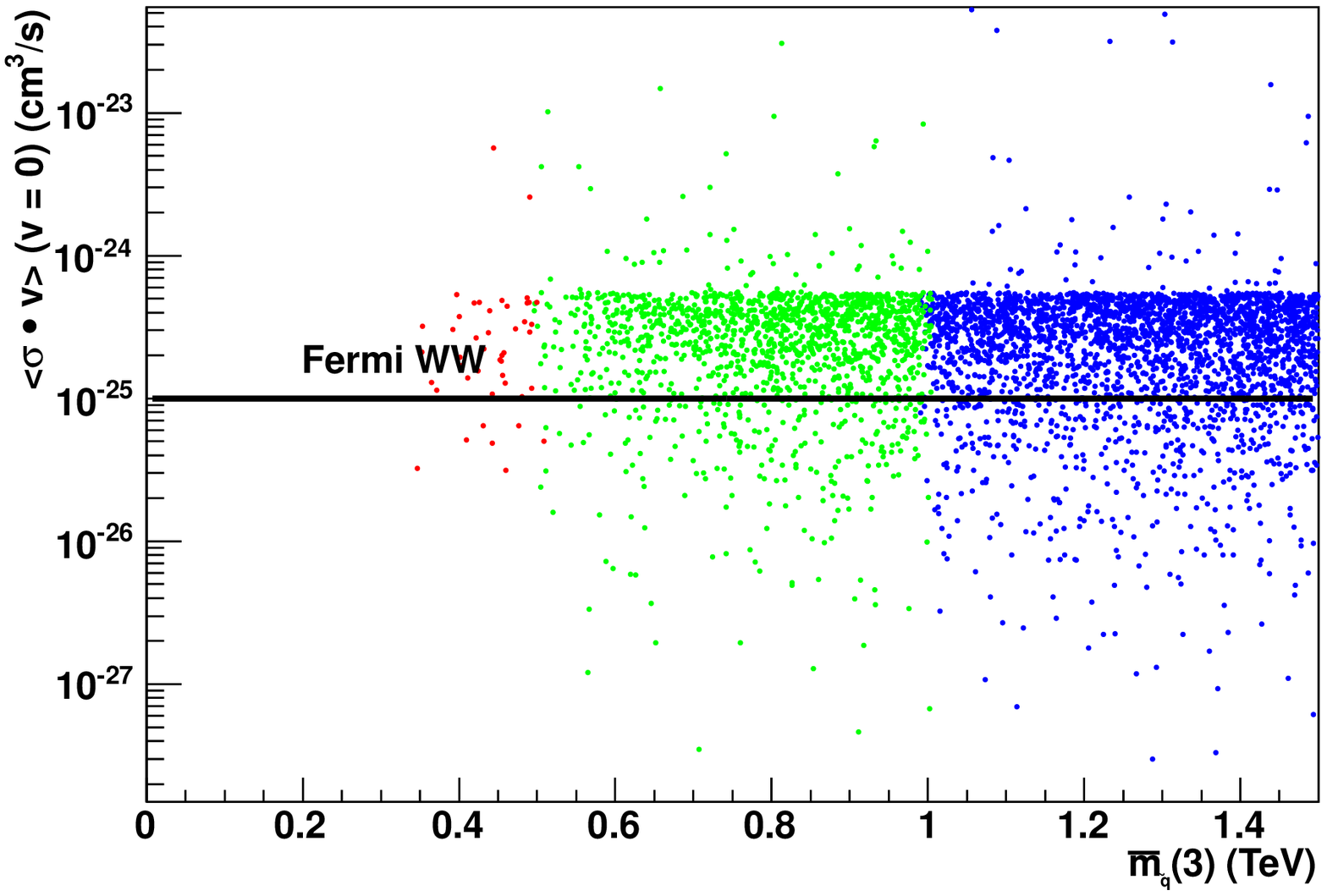}
\caption{Thermally averaged neutralino annihilation cross section times
relative velocity in limit as $v\to 0$ versus $m_0(3)$ for NS models
with $\mu =150$ GeV together with the bound from the Fermi satellite on
the cross section times velocity for WIMP annihilation to $W$-pairs
(assuming higgsino-like WIMPs saturate the
measured dark matter abundance).  }
\label{fig:sigv}}
%

\section{Summary and conclusions}
\label{sec:conclude}

The Natural SUSY model is defined by distinctive spectra characterized
by a low $|\mu| \sim m_h \alt 150-200$~GeV, with a rather light spectrum
of third generation squarks $m_{\tst_1},\ m_{\tst_2},\ m_{\tb_1}$~$ \alt
0.5-1.5$~TeV to stabilize the electroweak scale. In addition, $m_{\tg}\alt
4$ TeV so that loop corrections to third generation squark masses are
smaller than the squark mass.  First/second generation sfermions, on the
other hand, could be at the tens of TeV scale, thus suppressing unwanted
flavor-violating and $CP$-violating processes. Motivated by gauge
coupling unification, we expect the MSSM, or MSSM plus gauge singlets,
to be the effective field theory between $M_{\rm weak}$ and $M_{\rm
GUT}$.  In this case, the Natural SUSY mass spectra should arise
from underlying fundamental parameters that have their origin in GUT
scale physics. In
this paper, we determine the values of GUT scale parameters which lead
to models of natural SUSY. We find that, at the GUT scale, \bi
\item third generation mass parameters, $m_0(3)\sim 0.5-4$ TeV, 
\item first/second generation mass parameters, $m_0(1,2)\sim 5-25$ TeV, 
\item unified gaugino mass parameters, $m_{1/2}\sim 0.3-1.7$ TeV, and
\item the trilinear (third generation) scalar coupling,
$A_0/m_0(3)\agt-2$ \ei 
yield models with a natural SUSY
spectrum. The range of $\tan\beta$ and $m_A$ are relatively
unrestricted. Note that there is an upper bound on $m_0(1,2)$: values
much larger than about 25 TeV push third generation squarks into the
tachyonic range via 2-loop RGE effects.  We also find that values of
$m_h\sim 125$ GeV are very difficult to reconcile with a spectrum with
very light third generation scalars ($\msq3 <0.5$ TeV), but values of
$m_h$ up to 124~GeV can be realized if we allow $\msq3$ up to $1-1.5$ TeV
instead. Since some third generation squarks and charginos are rather light in
natural SUSY, the constraint from $BF(b\to s\gamma )$ is rather strong,
but models can be found which are consistent with the measured
branching fraction.  We provide some representative benchmark points for
low and high values of $m_h$.

At the LHC, the higgsino-like light charginos and neutralinos have only
small energy release in their decays, and so will be difficult to
observe, as noted for the related ``hidden SUSY'' scenario\cite{bbh}. 
However, in the case of natural SUSY, all four third generation squarks may be
produced at observable rates.  Sometimes, the lightest of these may have
just one decay mode accessible ({\it e.g.} case NS1 in this paper), and
thus may be described by an analysis using simplified models (this is
essentially impossible if $\tb_1$ is the lightest squark). However,
other cases arise where several different cascade decay possibilities
are open. The heavier third generation squarks decay via numerous modes,
and could lead to novel signatures involving $h$ or $Z$ from their cascade
decays.  Gluinos are favored to be rather heavy and frequently beyond
LHC reach, although cases where $m_{\tg}\alt 1.5$~TeV do occur
especially for $\msq3 \sim 1-1.5$~TeV.

At a linear $e^+e^-$ collider, we expect pair production of the
higgsino-like light charginos and neutralinos to offer a lucrative
discovery program of physics though specialized search strategies will
be needed to pull out the rather soft signal events. In addition, it is
possible that several third generation squarks and sleptons may be
accessible to a LC with $\sqrt{s}$ extending up to $\sim 1$ TeV or
beyond. Although these may decay to the light chargino as well as two
lighter neutralinos, it will be challenging to sort out the various
signals from the electroweak-ino cascades with very small secondary mass
gaps. To our knowledge there are no dedicated studies for event
topologies with this novel spectrum.

In Natural SUSY models, the higgsino-like neutralino $\tz_1$ is lightest
MSSM particle, and standard relic density calculations predict an
underabundance of higgsino-like WIMPS by a factor typically 15. Such an
under-abundance can be easily boosted to higher values if 1. there are late
decaying moduli fields with large branching fractions to
SUSY particles, or 2. if the PQ solution to the strong $CP$ problem is
invoked, whereupon thermal production of heavy axinos followed by
late-time decays in the early universe can augment the higgsino
abundance.  In this latter case, any remaining under-abundance can be
filled by axions. In the case of higgsino dominance of the dark matter
abundance, then we expect higgsino-like WIMPs to be detected by the next
round of direct and indirect WIMP detection experiments. An axion signal
could also be a viable possibility.

\acknowledgments

We thank Jenny List, Azar Mustafayev and Roman Nevzorov for discussions. 
This work was supported in part by the U.S. Department of Energy under grant Nos.~DE-FG02-04ER41305,
DE-FG02-04ER41291 and DE-FG02-95ER40896. 


%

\end{document}